\documentclass[aps,prd,preprint,superscriptaddress]{revtex4}
\usepackage{graphicx}
\usepackage{amsmath,amssymb}
\usepackage{bm}% bold math
\usepackage[caption=false]{subfig}
\usepackage{color}
\usepackage{setspace}
\usepackage{soul}
\usepackage{times}

\def\p{\partial}

\def\=:{=\hspace{-.7em}\raisebox{1.1ex}{.}\hspace{.1em}\raisebox{-0.2ex}{.} }

\newcommand {\beq}{\begin{eqnarray}}
\newcommand {\eeq}{\end{eqnarray}}
\newcommand {\non}{\nonumber\\}

\newcommand {\1}[1]{\frac{1}{#1}}

\newcommand {\ph}{\varphi}

\newcommand {\del}{\partial}

\newcommand {\tr}{{\rm tr}\,}

\begin{document}

% Use the \preprint command to place your local institutional report
% number in the upper righthand corner of the title page in preprint mode.
% Multiple \preprint commands are allowed.
% Use the 'preprintnumbers' class option to override journal defaults
% to display numbers if necessary

%Title of paper
\title{Fractional instantons and bions in 
the principal chiral model\\ 
on ${\mathbb R}^2\times S^1$ 
with twisted boundary conditions}

% repeat the \author .. \affiliation  etc. as needed
% \email, \thanks, \homepage, \altaffiliation all apply to the current
% author. Explanatory text should go in the []'s, actual e-mail
% address or url should go in the {}'s for \email and \homepage.
% Please use the appropriate macro foreach each type of information

% \affiliation command applies to all authors since the last
% \affiliation command. The \affiliation command should follow the
% other information
% \affiliation can be followed by \email, \homepage, \thanks as well.

\author{Muneto Nitta}
\affiliation{Department of Physics, and Research and Education Center for Natural 
Sciences, Keio University, Hiyoshi 4-1-1, Yokohama, Kanagawa 223-8521, Japan
}
%\homepage[]{Your web page}
%\thanks{}
%\altaffiliation{}

%Collaboration name if desired (requires use of superscriptaddress
%option in \documentclass). \noaffiliation is required (may also be
%used with the \author command).
%\collaboration can be followed by \email, \homepage, \thanks as well.
%\collaboration{}
%\noaffiliation

%Collaboration name if desired (requires use of superscriptaddress
%option in \documentclass). \noaffiliation is required (may also be
%used with the \author command).
%\collaboration can be followed by \email, \homepage, \thanks as well.
%\collaboration{}
%\noaffiliation

\date{\today}
\begin{abstract}
Bions are
multiple fractional instanton configurations 
with zero instanton charge 
playing important roles 
in quantum field theories on 
a compactified space  
with a twisted boundary condition. 
We classify fractional instantons and bions 
in the $SU(N)$ principal chiral model 
on ${\mathbb R}^2 \times S^1$ with 
twisted boundary conditions.
We find that fractional instantons
are global vortices wrapping around $S^1$
with their $U(1)$ moduli twisted along $S^1$,
that carry $1/N$ instanton (baryon) numbers 
for the ${\mathbb Z}_N$ symmetric twisted boundary condition
and irrational instanton numbers 
for generic  boundary condition. 
We work out neutral and charged bions 
for the $SU(3)$ case with 
the ${\mathbb Z}_3$ symmetric twisted boundary condition.
We also find for 
generic  boundary conditions  
that only the simplest neutral bions 
have zero instanton charges
but instanton charges are not canceled out for 
charged bions.
A correspondence between 
fractional instantons and bions 
in the $SU(N)$ principal chiral model 
and those in Yang-Mills theory is given  
through a non-Abelian Josephson junction.

\end{abstract}

% insert suggested PACS numbers in braces on next line
% insert suggested keywords - APS authors don't need to do this
%\keywords{}

%\maketitle must follow title, authors, abstract, \pacs, and \keywords
\maketitle

\section{Introduction}

Recently, 
much attention have been paid to  
quantum field theories on compactified spaces 
${\mathbb R}^d \times S^1$ 
with twisted boundary conditions, 
such as QCD with adjoint fermions 
on ${\mathbb R}^3 \times S^1$
and nonlinear sigma models on 
${\mathbb R}^1 \times S^1$, 
that admit  fractional instantons and 
 bions, {\it i.~e.~}
multiple fractional instanton configurations 
with vanishing instanton charge 
\cite{Unsal:2007vu}--\cite{Shermer:2014wxa}. 
Magnetic (charged) bions 
carry a magnetic charge
and are conjectured to lead semiclassical confinement 
in QCD on ${\mathbb R}^{3} \times S^{1}$ 
\cite{Anber:2011de},
while neutral bions carry no magnetic charge 
and are identified as the infrared renormalons
\cite{Argyres:2012vv}--\cite{Cherman:2014ofa}
(see Refs.~\cite{'tHooft:1977am, 
Fateev:1994ai, Fateev:1994dp} for earlier works) 
playing an essential role in 
self-consistent 
semiclassical definition of quantum field theories 
through the resurgence \cite{Ec1}. 

In lower dimensions, 
fractional instantons were found 
in the ${\mathbb C}P^{N-1}$ model 
\cite{Eto:2004rz,Eto:2006pg} 
(see also Refs.~\cite{Bruckmann:2007zh} for 
subsequent study)
and the Grassmann sigma model \cite{Eto:2006mz} 
on ${\mathbb R}^1 \times S^1$ 
with twisted boundary conditions. 
Bions and their role in the resurgence 
have been extensively studied in 
the ${\mathbb C}P^{N-1}$ model
\cite{Dunne:2012ae, Dunne:2012zk, Dabrowski:2013kba,
Misumi:2014jua,Shermer:2014wxa}
and the Grassmann sigma model 
\cite{Misumi:2014bsa,Dunne:2015ywa}
on ${\mathbb R}^1 \times S^1$. 
The former admits only neutral bions while the 
latter admits both neutral and charged bions \cite{Misumi:2014bsa}.
Fractional instantons and bions 
in the $O(N)$ nonlinear sigma model 
on ${\mathbb R}^{N-2} \times S^1$ 
have been studied recently 
with general twisted boundary conditions 
in which arbitrary number of fields changes signs 
\cite{Nitta:2014vpa}.
The $O(3)$ model is equivalent to 
the ${\mathbb C}P^1$ model 
studied before \cite{Dunne:2012ae,Dunne:2012zk,Dabrowski:2013kba,
Bolognesi:2013tya,Misumi:2014jua}.
The $O(4)$ model is equivalent to a principal chiral model 
with a group $SU(2)$ 
(or a Skyrme model if four derivative term is added \cite{Skyrme:1962vh}), 
for which the case of the boundary condition 
with two fields changing their signs  
is equivalent to the 
${\mathbb Z}_2$ (center) symmetric boundary condition.
In this case, fractional instantons 
are vortices winding around $S^1$ 
with $U(1)$ moduli twisted half along $S^1$.

In this paper, 
we study the $SU(N)$ principal chiral model 
on ${\mathbb R}^2 \times S^1$ with 
twisted boundary conditions. 
Previously the principal chiral models were studied 
in two dimensions \cite{Cherman:2013yfa,Cherman:2014ofa} 
for which instantons do not exist. 
We study the principal chiral model 
in three dimensions, where instantons exist with 
the instanton number defined by  
the third homotopy group $\pi_3$ 
that is also known as baryon number 
in the context of the Skyrme model \cite{Skyrme:1962vh}. 
We show that this case allows 
$N-1$ kinds  of global vortices 
accompanied by $U(1)$ moduli,  
and fractional instantons
are vortices   
wrapping around the $S^1$ direction, 
with $U(1)$ moduli twisted along $S^1$ 
by the angle $2\pi/N$ (or its complement) 
for the ${\mathbb Z}_N$ center 
symmetric twisted boundary condition
and by generic angle for 
generic  boundary conditions. 
They carry $1/N$ instanton (baryon) numbers
for the ${\mathbb Z}_N$ symmetric twisted boundary 
condition and irrational instanton numbers 
for generic  boundary condition.
We classify neutral and charged bions 
for the $SU(3)$ case with 
the ${\mathbb Z}_3$ symmetric twisted boundary condition.
We also point out that 
for the cases with generic boundary conditions  
only the simplest neutral bions, 
composed of a set of 
a fractional instanton and fractional anti-instanton, 
have zero instanton charges  
but instanton charges are not canceled out for 
charged bions. 
We further discuss 
a correspondence between 
fractional instantons and bions 
in the $SU(N)$ principal chiral model 
and those in Yang-Mills theory;
the latter become the former 
if reside inside a non-Abelian domain wall 
\cite{Shifman:2003uh,Eto:2005cc,Eto:2008dm} 
(non-Abelian Josephson junction \cite{Nitta:2015mma}) 
in the Higgs phase \cite{Nitta:2015mxa}.

This paper is organized as follows.
In Sec.~\ref{sec:model}, we first give the 
$SU(N)$ principal chiral model.
In Sec.~\ref{sec:SU(2)} we review  
fractional instantons and bions 
in the $SU(2)$ principal chiral model 
on  ${\mathbb R}^2 \times S^1$ with 
the center symmetric twisted boundary conditions.
We find charged bions that were not studied before.
In Sec.~\ref{sec:SU(3)} we work out   
fractional instantons and bions 
in the $SU(3)$ principal chiral model 
on  ${\mathbb R}^2 \times S^1$ with 
the center symmetric twisted boundary conditions. 
In Sec.~\ref{sec:SU(N)}, we briefly 
discuss  the $SU(N)$ principal chiral model.
In Sec.~\ref{sec:general}, generic boundary 
conditions are discussed for $SU(N)$.
In Sec.~\ref{sec:YM}, we discuss 
the relation between fractional instanton and bions 
in the $SU(N)$ principal chiral model 
and those in the $SU(N)$ Yang-Mills theory.
Sec.~\ref{sec:summary} is devoted to a summary 
and discussion.

%%%%%%%%%%%%%%%%%%%%%%%%%%
\section{The principal chiral model on ${\mathbb R}^2 \times S^1$ with twisted boundary conditions
\label{sec:model}}

Let  $U(x)$ be scalar fields taking a 
value in the group $G=SU(N)$. 
Then, the Lagrangian of the $SU(N)$ principal chiral model 
is given as
\beq
\mathcal{L} = 
f_{\pi}^2 \tr (\p_{\mu}U^{\dagger} \p^{\mu} U) 
\eeq
with a decay constant $f_{\pi}$.
The symmetry of the Lagrangian 
 is 
$G = SU(N)_{\rm L} \times SU(N)_{\rm R}$ 
($U \to U'= g_{\rm L} U g_{\rm R}^\dag$),
that is spontaneously broken down to 
$H \simeq SU(N)_{\rm L+R}$  
($U \to U'= g U g^\dag$). 
The target space is 
$M \equiv G/H \simeq SU(N) \ni U(x)$. 
The instanton number $B$ in $d=3+0$ dimensions
(or equivalently the baryon number or Skyrme charge 
in $d=3+1$ dimensions),
 taking a value in the third homotopy group 
$B \in \pi_3(M)$, 
is defined as $(i=1,2,3)$
\beq
B &=& -\1{24\pi^2} \int d^3x \; \epsilon^{ijk} 
\tr \left( U^\dag\p_i U U^\dag\p_j U U^\dag\p_k U\right) \non
&=& \1{24\pi^2} \int d^3x \; \epsilon^{ijk} 
\tr \left( U^\dag\p_i U\p_j U^\dag\p_k U\right) .
\eeq

We consider the space ${\mathbb R}^2 \times S^1$ 
with non-trivially twisted boundary conditions along $S^1$.
The ${\mathbb Z}_N$ symmetric twisted boundary condition 
for the $SU(N)$ principal chiral model is defined by
\beq 
&& U(x^1,x^2,x^3+R)=  W U(x^1,x^2,x^3) W^\dagger, \,\,\non
&& W = {\rm diag.} (1,\omega,\omega^2,\cdots,\omega^{N-1}) 
=  \exp\left[ {2\pi i \over N}  {\rm diag.} (0,1,\cdots,N-1)\right], 
\quad
\omega = e^{2\pi i \over N} .
\label{eq:tbcSU(N)}
\eeq
The ${\mathbb Z}_2$ twisted boundary condition 
for the $SU(2)$ case is
\beq 
\, U(x^1,x^2,x^3+R) \,=\, 
 W U(x^1,x^2,x^3) W^\dagger, \quad 
 \,  W \,=\, \sigma_3 \,=\, {\rm diag.} (1,-1). \, \label{eq:tbcSU(2)}
\eeq
The $SU(2)$ principal chiral model
is equivalent to the $O(4)$ nonlinear sigma model.
If we define four real scalar fields $n_A(x)$  ($A=1,2,3,4$) 
from the 
$SU(2)$-valued field $U(x)$ by
\beq
U = i \sum_{a=1,2,3} n_a \sigma_a + n_4 \mathbf{1}_2 
\eeq
with  the Pauli matrices  $\sigma_a$ and 
the constraint $\mathbf{n}\cdot\mathbf{n}=1$
equivalent to $U^\dag U = \mathbf{1}_2$,
the boundary condition 
(\ref{eq:tbcSU(2)})
becomes $(n_1,n_2,n_3,n_4)(x^1,x^2,x^3+R) = (-n_1,-n_2,n_3,n_4)(x^1,x^2,x^3)$ 
that we called $(-,-,+,+)$ \cite{Nitta:2014vpa}.

We also consider more general twisted 
boundary condition 
\beq
&& U(x^1,x^2,x^3+R) \,=\,  W U(x^1,x^2,x^3) W^\dagger, \non
&& W \,=\, {\rm diag.} (e^{i m_1},e^{i m_2},\cdots,e^{im_N}) ,
\quad m_1 \leq m_2 \leq \cdots \leq m_N.
\label{eq:tbcSU(N)2}
\eeq
In this paper, we first focus on 
the ${\mathbb Z}_N$ symmetric twisted boundary condition 
in Eq.~(\ref{eq:tbcSU(N)}),
that corresponds to $m_a= 2\pi (a-1) /N$. 
We also consider the generic non-degenerate case later.

In small compactification radius limit, 
the Scherk-Schwarz dimensional reduction is 
effectively induced, to yield a potential term 
(twisted mass) \cite{Nitta:2015mma}
\beq
&& V = 
  f_{\pi}^2  \tr ([ M, U ]^\dagger[M, U ])  \label{eq:twisted-mass}
\eeq
with $M \equiv (m_1,m_2,\cdots,m_N)$.

%%%%%%%%%%%%%%%%%%%%%%%%%%%%%%%%
\section{Fractional instantons and bions in the 
$SU(2)$ principal chiral model}
\label{sec:SU(2)}

%%%%%%%%%%%%%%%%%%%

In this section, we consider the $SU(2)$ principal chiral model 
with the ${\mathbb Z}_2$ symmetric boundary condition.
This section is mostly rewriting the results in Ref.~\cite{Nitta:2014vpa} in terms of the principal chiral field $U(x)$ 
because 
the $SU(2)$ principal chiral model is equivalent to 
the $O(4)$ sigma model, 
but we will find it useful for warming up 
to study the $SU(N)$ principal chiral model.  
Charged bions in the third subsection is a new result 
that  were not studied before.

\subsection{Fractional instantons}\label{sec:SU(2)fractional}

%%%%%%%%%%%%%%%%%%%%%
\begin{figure}
\begin{center}
\begin{tabular}{cccc}

$(+1;+\1{2})$ & 
$(-1;+\1{2})$ &
$(-1;-\1{2})$ &  
$(+1;-\1{2})$\\ \hline
\includegraphics[width=0.19\linewidth,keepaspectratio]{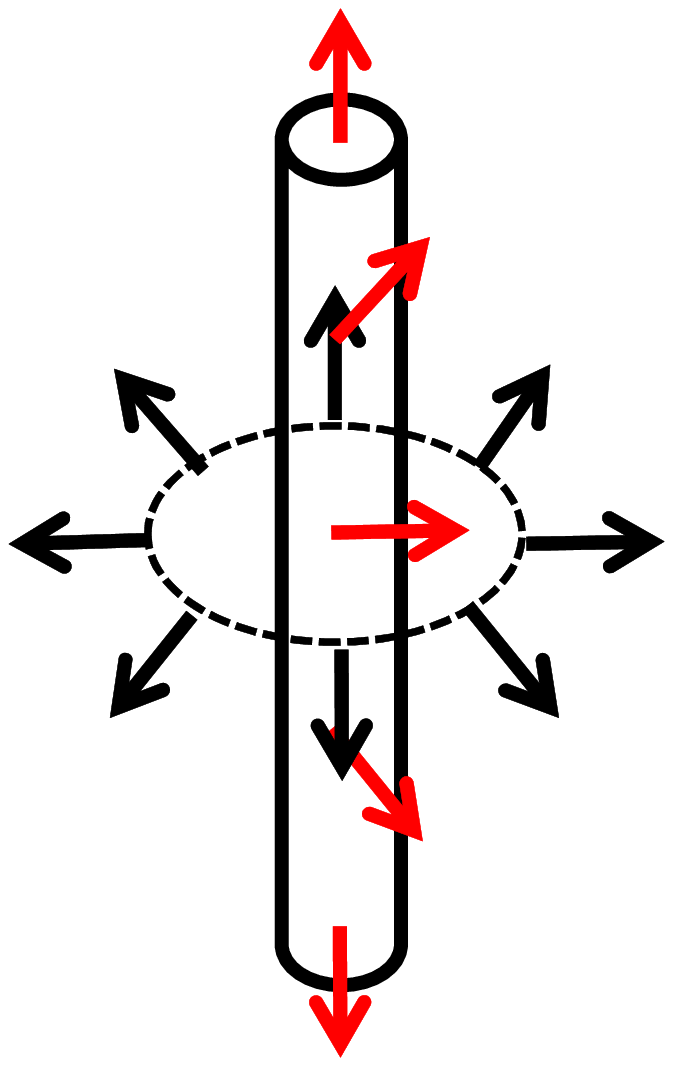} &
\includegraphics[width=0.094\linewidth,keepaspectratio]{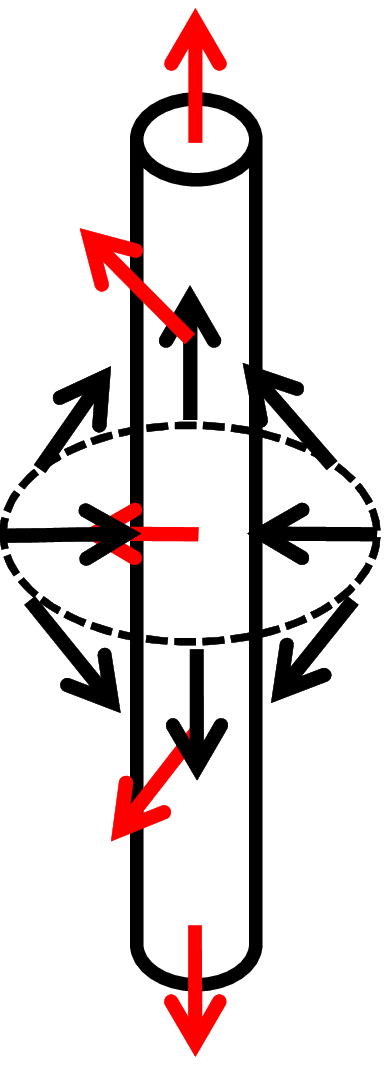} &
\includegraphics[width=0.094\linewidth,keepaspectratio]{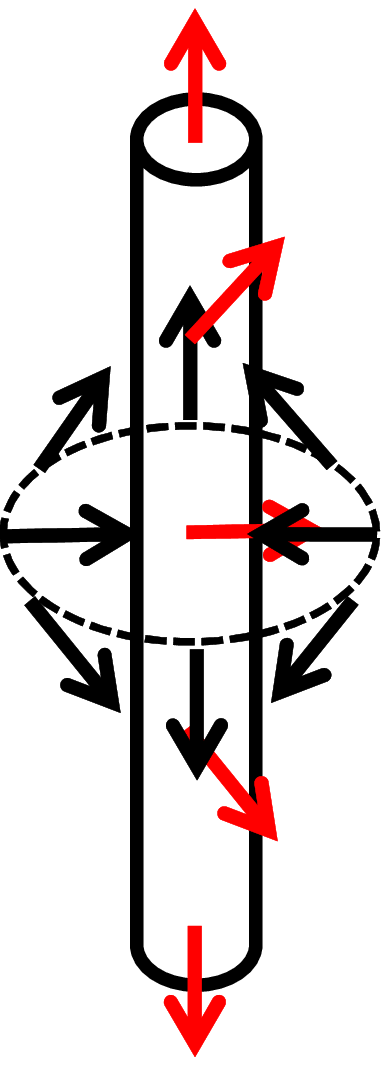} &
\includegraphics[width=0.19\linewidth,keepaspectratio]{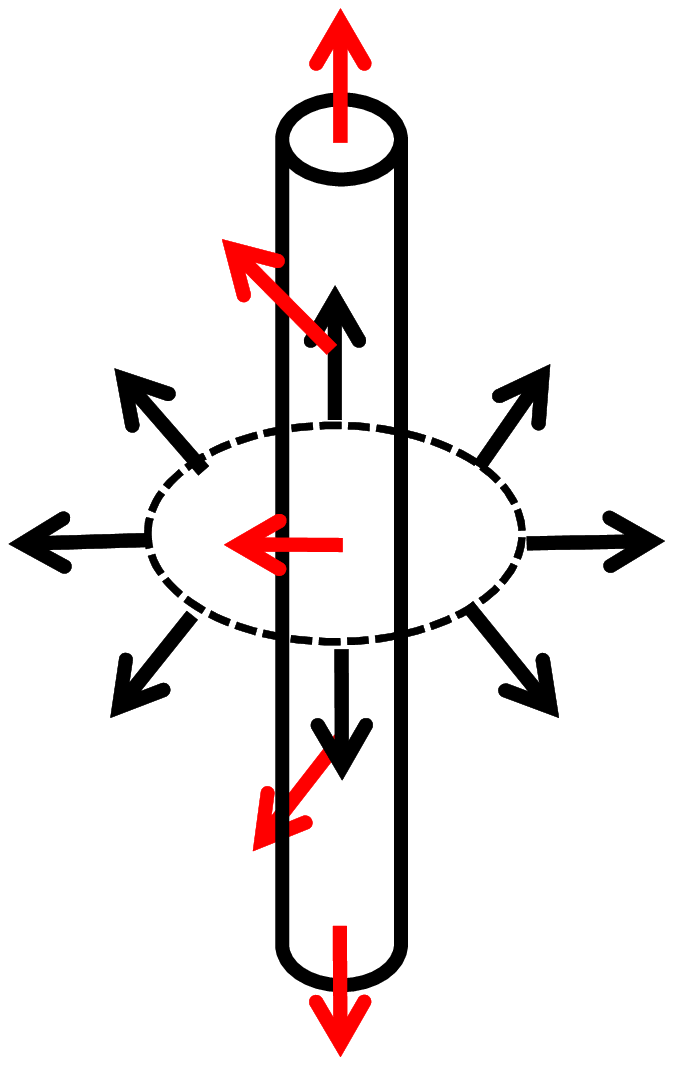} \\ 
(a) & (b) & (c) & (d)
\end{tabular}
\end{center}
\caption{
Fractional instantons in the $SU(2)$ 
principal chiral model 
(figures are taken from Ref.~\cite{Nitta:2014vpa}).
The first lines indicate 
the topological charges (homotopy groups) 
$(\pi_{1}({\cal N});\pi_3(M))$ for the vortices 
and instantons (Skyrmions). 
The black arrows denote the $U(1)$ moduli of the vacua 
while the red arrows denote the $U(1)$ moduli of the vortices. 
Fractional (anti-)instantons can constitute 
following composite structures:   
(a)+(b) instanton, (c)+(d) anti-instanton, 
(a)+(c), (b)+(d) neutral bions,
(a)+(d), (b)+(c) charged bions. 
\label{fig:SU(2)}
}
\end{figure}
%%%%%%%%%%%%%%%%%%%%%%

%%%%%%%%%%%%%%%%%%%%%
\begin{figure}
\begin{center}
\includegraphics[width=0.09\linewidth,keepaspectratio]{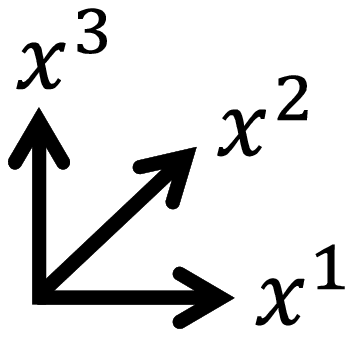} 
\includegraphics[width=0.90\linewidth,keepaspectratio]{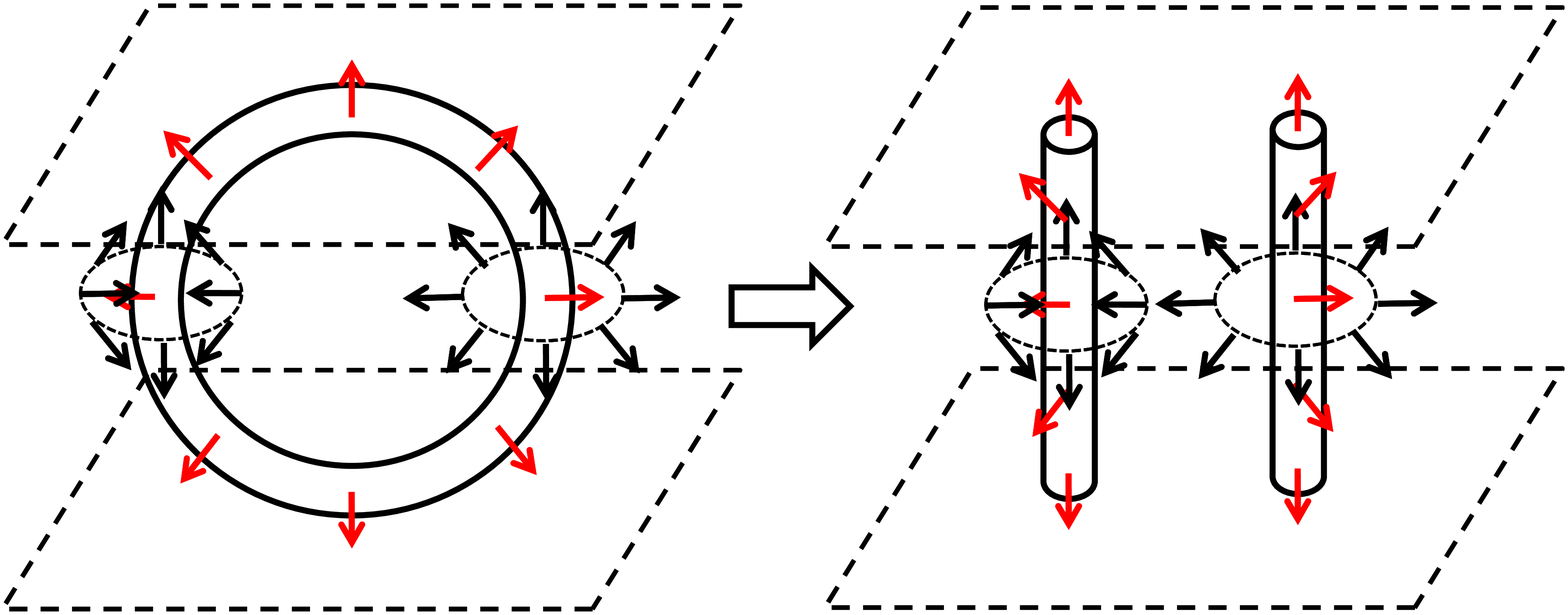} 
\end{center}
\caption{Decay of a twisted closed vortex string of the size of the compact direction into two fractional 
instantons
in the $SU(2)$ principal chiral model 
(figures are taken from Ref.~\cite{Nitta:2014vpa}).
The notations are the same with Fig.~\ref{fig:SU(2)}.
The dotted planes denote the boundary at $z=0$ and $z=R$ 
where the fields are twisted. 
When a closed vortex touches to itself through 
the compact direction $z$, 
a reconnection of the two parts of the string occurs 
to be split into two fractional (anti-)instantons, 
that is, vortices winding around $S^1$ with the half twisted $U(1)$ moduli.
\label{fig:decay-ring-O4}}
\end{figure}
%%%%%%%%%%%%%%%%%%%%%%%

Fractional instantons in the principal chiral model 
were classified into four kinds, 
as illustrated in Fig.~\ref{fig:SU(2)}.
These can be obtained as follows.
The fixed manifold ${\cal N}$ under the action 
that acts on the boundary condition is
\beq
 {\cal N} \simeq U(1) \simeq S^1
\eeq
that is generated by $\sigma_3$.
Therefore, it has a nontrivial first homotopy group 
\beq
 \pi_1(S^1) \simeq {\mathbb Z}.
\eeq 
Let us place a vortex along the $z=x^3$ direction.
The ansatz for a vortex configuration 
can be given as
\beq
U(r,\theta,z) = 
\left(\begin{array}{cc} 
   \cos f(r) e^{+i \theta} & - \sin f (r) e^{+i\alpha(z)} \\
   \sin f(r) e^{-i\alpha(z)} &   \cos f(r) e^{-i \theta} 
\end{array}
\right), \label{eq:SU(2)-vortex2}
\eeq
where $(r,\theta,z)$ are cylindrical coordinates 
$f$ is a profile function satisfying 
the boundary conditions
\beq
  f \to 0 
\; {\rm for}\; r\to \infty, \quad
  f = \pi/2 
\; {\rm for}\; r=0. \label{eq:bc-f}
\eeq
An anti-vortex can be obtained as
$U(r,-\theta,z)$. 
In Eq.~(\ref{eq:SU(2)-vortex2}),  
$\alpha$ 
is a $U(1)$ modulus of the vortex,
that is constant if the vortex does not 
wind around $S^1$.  
When the vortex winds around
$S^1$ with the twisted boundary condition in
 Eq.~(\ref{eq:tbcSU(2)}), 
the modulus $\alpha$ has to satisfy 
the boundary condition
\beq
 \alpha(z+R) =  \alpha(z) \pm \pi .\label{eq:tbc-alpha-SU2}
\eeq
The following $z$-dependence of $\alpha$ satisfy the boundary condition;
\beq
 \alpha (z) = \alpha_0 \pm {\pi \over R} z 
\eeq
that we denote $\alpha^+$ and $\alpha^-$, respectively.

The topological instanton charge 
(baryon number) 
can be calculated as
\begin{align}
B = \frac{1}{16\pi^2} \int d^3x \; \frac{1}{r}\sin(f) f_r \alpha_z 
= \pm \1{2\pi} \int_0^R dz \del_z \alpha
= \pm \1{2\pi} [\alpha]^{z=R}_{z=0} 
= 
\Bigg\{\begin{array}{c}
\pm \displaystyle{{1\over 2}}      
   \quad {\rm for}\;\; \alpha=\alpha^+\\
\mp \displaystyle{{1 \over 2}}  
   \quad {\rm for}\;\; \alpha=\alpha^-
\end{array} ,
\end{align}
where the upper and lower signs correspond to
 a vortex and anti-vortex, respectively.
More generally, a vortex string with the winding number $Q$, 
along which  the $U(1)$ modulus is twisted $P$ 
times, 
has 
the instanton number  
 $B=PQ$ \cite{Gudnason:2014hsa}
(which was obtained in Ref.~\cite{Kobayashi:2013aza}  
to calculate the Hopf number for Hopfions 
by lifting up $\pi_3(S^2)$ to $\pi_3(S^3)$).
The topological charges of fractional (anti-)instantons 
with the ${\mathbb Z}_2$ symmetric 
twisted boundary condition are summarized in 
Table \ref{table:homotopy-O4-2}.
%%%%%%%%%%%%%%%%%%%%%%%%%
\begin{table}[h]
\begin{tabular}{c|c|c|c|c} 
$(v;B)$ & $\pi_1({\cal N})$ & $\pi_1$ (${\cal M})$& $B\in \pi_3(M)$ & Fig\\ \hline
$(+1;+1/2)$ &$ +1$     & $+1/2$ & $+1/2$ & Fig.~\ref{fig:SU(2)} (a) \\
 $(-1;+1/2)$   &$ -1$     & $-1/2$ & $+1/2$ & Fig.~\ref{fig:SU(2)} (b)  \\
$(-1;-1/2)$  &$ -1$     & $+1/2$ & $-1/2$ & Fig.~\ref{fig:SU(2)} (c)  \\
$(+1;-1/2)$  &$ +1$     & $-1/2$ & $-1/2$ & Fig.~\ref{fig:SU(2)} (d) 
\end{tabular}
\caption{Homotopy groups of fractional (anti-)instantons in the 
$SU(2)$ principal chiral model 
with the ${\mathbb Z}_2$ symmetric twisted boundary condition.
The columns represent the homotopy groups  
of a vortex $\pi_1$, a $U(1)$ modulus $\pi_1$, 
and the instanton $\pi_3$ from left to right. 
\label{table:homotopy-O4-2}}
\end{table}
%%%%%%%%%%%%%%%%%%%%%%

It is known that
a single instanton (Skyrmion) can be represented by 
(a global analog of) a vorton \cite{Davis:1988jq}, 
{\it i.~e.~}, a closed vortex string along which a $U(1)$ modulus 
is twisted \cite{Ruostekoski:2001fc,Nitta:2012hy},
which was first found in the context of 
Bose-Einstein condensates 
(see also \cite{Metlitski:2003gj}), 
and stable solutions in a Skyrme model were also 
constructed in Refs.~\cite{Gudnason:2014hsa,Gudnason:2014gla,
Gudnason:2014jga}.
A single instanton as a vorton is shown
 in the left panel in Fig.~\ref{fig:decay-ring-O4}. 
When the size of the closed vortex string is of 
the same with 
that of the compactification scale $R$, 
the closed vortex string touches itself 
through the compact $x^3$ direction with 
the twisted boundary condition.
Subsequently 
a reconnection of two fractions of the closed string 
occurs 
(see Ref.~\cite{Eto:2006db} for a reconnection of strings with moduli). 
Then, 
the closed string is split 
into two vortex strings winding around the compact direction,  
and subsequently 
they are separated into the $x$-$y$ plane, 
as illustrated in the right panel of Fig.~\ref{fig:decay-ring-O4}.
The $U(1)$ modulus is twisted half along 
each string, resulting in a fractional (anti-)instanton. 
We thus find four kinds of fractional (anti-)instantons, 
as summarized in Fig.~\ref{fig:SU(2)}  (a)--(d).

The Skyrme term is not  needed for the stability
even though fractional instantons are Skyrmions 
as was demonstrated in Ref.~\cite{Gudnason:2014hsa},
in which  
stable configurations 
of (half) Skyrmions inside a vortex string 
were constructed without the Skyrme term 
(on ${\mathbb R}^3$ 
without twisted boundary condition).

All vortices are global vortices having 
the divergent energy
\beq
  E \sim \log \Lambda /\xi
\eeq
at large distance, apart from finite contribution 
from the core.
Here, $\Lambda$ is the system size  in the $x$-$y$ plane, 
and $\xi \sim m^{-1}$ is the size of the vortex core.

Fractional instantons 
are global vortices in the $x$-$y$ plane
so that the interaction between them 
is 
\beq 
  E_{\rm int} \sim \mp \log \rho/\xi, 
\quad F_{\rm int} = - {\del E_{\rm int} \over \del\rho} 
 \sim \pm 1/\rho
\eeq
with distance $\rho$ for large separation 
$\rho \gg \xi$,  
where the upper signs are for a pair of (anti-)vortices 
(repulsion)
and the lower signs are for a pair of a vortex and anti-vortex 
(attraction).
The interaction between two vortices 
at short distance $\rho \sim \xi$ 
depends on the moduli $\alpha$ 
in the cores, but 
we do not discuss it in this paper.

%%%%%%%%%%%%%%%%%%%%

\subsection{Neutral bions}\label{sec:SU(2)neutral}
Neutral bions in the $SU(2)$ principal chiral model 
were discussed before in 
Ref.~\cite{Nitta:2014vpa}.
Neutral bions  are configurations with 
zero instanton charges and zero vortex charges:
\beq
 \sum_i (v_i;B_i) = (0;0) . \label{eq:neutral}
\eeq
Neutral bions composed of two fractional 
(anti-)instantons can be constructed from 
 fractional instantons with 
the {\it opposite} vortex charges with the 
{\it same} winding  of the $U(1)$ modulus 
along $z$, 
that is, a configuration 
composed of (a) and (c) 
or  (b) and (d) in Fig.~\ref{fig:SU(2)}.

The interaction between fractional instantons 
constituting a neutral bion is attractive, 
because they are a pair of a global vortex 
and global anti-vortex:
\beq 
 E_{\rm int} \sim + \log \rho/\xi, \quad
 F_{\rm int} \sim - 1/ \rho
\eeq 
with distance $\rho$ for large separation.

%%%%%%%%%%
\subsection{Charged bions}\label{sec:SU(2)charged}
Charged bions were not discussed before 
in the $SU(2)$ principal chiral model.
Charged bions  are configurations with 
zero instanton charges and non-zero vortex charges:
\beq
 \sum_i (v_i;B_i) = (v;0), \quad v\neq 0 . \label{eq:charged}
\eeq
For charged bions composed of two fractional 
(anti-)instantons, one prepares
fractional instantons with 
the {\it same} vortex charges with the 
{\it opposite} winding  of the $U(1)$ modulus 
along $z$,
that is, the configurations 
$(v;B)=(2;0)$ for (a) and (d) in Fig.~\ref{fig:SU(2)},
$(v;B)=(-2;0)$ for  (b) and (c) in Fig.~\ref{fig:SU(2)}.

The interaction between fractional instantons 
constituting a charged bion is  repulsive 
because they are a pair of  global vortices:
\beq 
 E_{\rm int} \sim - \log\rho/\xi, \quad 
 F_{\rm int} \sim + 1/\rho
\eeq 
with distance $\rho$ for large separation.

%%%%%%%%%%%%%%%%%%%%%%%%%%%%%%%%%
\section{Fractional instantons and bions in the 
$SU(3)$ principal chiral model}\label{sec:SU(3)}

\subsection{Fractional instantons}\label{sec:SU(3)fractional}

We consider the ${\mathbb Z}_3$ symmetric 
twisted boundary condition:
\beq
 W = {\rm diag.} (1,\omega,\omega^2)
=  \exp\left[ {2\pi i \over 3}  {\rm diag.} (0,1,2)\right], 
\quad
\omega = e^{2\pi i \over 3} .
\eeq
The fixed manifold ${\cal N}$ is 
\beq
&& U = {\rm diag.} (e^{i\alpha},e^{i\beta},e^{-i\alpha-i\beta}),
\quad
{\cal N} \simeq U(1)^2. \label{eq:homotopy}
\eeq
The non-trivial first homotopy group 
\beq
 \pi_1 ({\cal N}) \simeq {\mathbb Z} \oplus  {\mathbb Z} 
\eeq
admits homotopically distinct two kinds of vortices. 
The fundamental, {\it i.~e.,~} minimum winding vortices in the $SU(3)$ principal chiral model can be obtained by embedding 
the one in Eq.~(\ref{eq:SU(2)-vortex2}) of 
the $SU(2)$ principal chiral model to the $SU(3)$ matrix:
\beq
&& U_1 (r,\theta,z) = 
\left(\begin{array}{ccc} 
   \cos f(r) e^{+i \theta} & - \sin f (r) e^{+i\alpha(z)} & 0\\
   \sin f(r) e^{-i\alpha(z)} &   \cos f(r) e^{-i \theta} & 0\\
   0 & 0 & 1
\end{array}
\right),\non
&& U_2 (r,\theta,z) = 
\left(\begin{array}{ccc} 
1 & 0 & 0\\
0 & \cos f(r) e^{+i \theta} & - \sin f (r) e^{+i\alpha(z)} \\
0 & \sin f(r) e^{-i\alpha(z)} &   \cos f(r) e^{-i \theta} 
\end{array}
\right),\non
&& U_3 (r,\theta,z) = 
\left(\begin{array}{ccc} 
   \cos f(r) e^{-i \theta} & 0 & + \sin f (r) e^{-i\alpha(z)} \\
 0 & 1 & 0\\
 -\sin f(r) e^{+i\alpha(z)} &  0 & \cos f(r) e^{+i \theta} 
\end{array}
\right).  \label{eq:SU(3)-vortex1} 
\eeq
Anti-vortices can be obtained as
$U_a(r,-\theta,z)$ with $a=1,2,3$.
These three are not homotopically 
independent of each other.  
We take the first and second as 
the independent basis of the first homotopy group 
in Eq.~(\ref{eq:homotopy}), 
in which 
the topological vortex charges in 
Eq.~(\ref{eq:SU(3)-vortex1}) are
\beq
 u_1=(1,0), \quad u_2=(0,1), \quad u_3=(-1,-1) \quad \in 
 {\mathbb Z} \oplus  {\mathbb Z}  \simeq \pi_1 ({\cal N}) ,
\eeq
respectively. 
The third one is not independent of the rests  
as can be seen from the fact that 
the vortex charges are canceled out 
when all three are present together:
\beq
 \sum_{a=1,2,3}u_a = (0,0).
\eeq
In Eq.~(\ref{eq:SU(3)-vortex1}), 
$\alpha$ is a $U(1)$ modulus of a vortex
that is constant if the vortex does not 
wrap the $S^1$ direction.

When a vortex wraps the $S^1$ direction, 
$\alpha$ must change along the vortex world-line
due to 
the twisted boundary condition in Eq.~(\ref{eq:tbcSU(N)}):
\beq
 \exp {i\alpha (z+R) } 
=  \exp {\left[i \alpha (z) + {2\pi i\over 3}\right]} ,
  \label{eq:tbc-phase}
\eeq
where $z$ is the coordinate along $S^1$ with the period 
$R$.
This boundary condition can be satisfied by 
the two different  minimum paths with 
the following $z$-dependence of $\alpha$: 
\beq
\begin{displaystyle}
 \alpha (z) = 
\Bigg\{\begin{array}{c}
\alpha^+(z) = \alpha_0 +  \displaystyle{{2\pi\over 3} {z\over R}} \\
\alpha^-(z) = \alpha_0 -  \displaystyle{{4\pi \over 3} {z\over R}}
\end{array} 
\end{displaystyle}
,  \label{eq:z-dep}
\eeq
with a constant $\alpha_0$. 
Correspondingly, 
each of them carries fractional instanton (baryon) number:
\begin{align}
B \,
= \, \pm \1{2\pi} [\alpha]^{z=R}_{z=0} =
\Bigg\{\begin{array}{c}
\pm \displaystyle{{1\over 3}}      
   \quad {\rm for}\;\; \alpha=\alpha^+\\
\mp \displaystyle{{2 \over 3}}  
   \quad {\rm for}\;\; \alpha=\alpha^-
\end{array} ,
\end{align}
where the upper and lower signs correspond to
 a vortex and anti-vortex, respectively.
We thus have found  
six (four independent) types of elementary 
fractional instantons 
as well as six  (four independent) types of 
elementary fractional anti-instantons, 
as summarized in Table \ref{table:classification-fractional}.
%%%%%%%%%%%%%%%%%%%%%%%%%
\begin{table}[h]
\begin{tabular}{c|cc|cc|c} 
  $(v_1,v_2;B)$ & $\pi_1({\cal N})$ & $U$ & $\pi_1({\cal M})$& $\alpha$    & $B\in \pi_3(M)$ \\ \hline
 $(+1,0;+1/3)$ & $(+1,0)$  & $U_1(\theta)$ & $+1/3$ & $\alpha^+$ & 
$+1/3$\\
 $(0,+1;+1/3)$ & $(0,+1)$  & $U_2(\theta)$ & $+1/3$ & $\alpha^+$ & $+1/3$\\
 $(-1,-1;+1/3)$ & $(-1,-1)$  & $U_3(\theta)$ & $+1/3$ & $\alpha^+$ & $+1/3$\\ 
\hline
 $(-1,0;+2/3)$ & $(-1,0)$  & $U_1(-\theta)$ & $-2/3$ &$\alpha^-$ & $+2/3$\\
 $(0,-1;+2/3)$ & $(0,-1)$  & $U_2(-\theta)$ & $-2/3$ &$\alpha^-$ & $+2/3$\\ 
 $(+1,+1;+2/3)$ & $(+1,+1)$  & $U_3(-\theta)$ & $-2/3$ &$\alpha^-$ & $+2/3$\\
\hline
 $(-1,0;-1/3)$ & $(-1,0)$  & $U_1(-\theta)$ & $+1/3$ &$\alpha^+$ & 
$-1/3$\\
 $(0,-1;-1/3)$ & $(0,-1)$  & $U_2(-\theta)$ & $+1/3$ &$\alpha^+$ & $-1/3$\\
 $(+1,+1;-1/3)$ & $(+1,+1)$  & $U_3(-\theta)$ & $+1/3$ &$\alpha^+$ & $-1/3$ \\ 
\hline
 $(+1,0;-2/3)$ & $(+1,0)$  & $U_1(\theta)$ & $-2/3$ &
$\alpha^-$ & 
$-2/3$\\
 $(0,+1;-2/3)$ & $(0,+1)$  & $U_2(\theta)$ & $-2/3$ &$\alpha^-$ & $-2/3$\\
 $(-1,-1;-2/3)$ & $(-1,-1)$  & $U_3(\theta)$ & $-2/3$ & $\alpha^-$ & $-2/3$\\ 
\end{tabular}
\caption{
We label configurations by $(v_1,v_2;B) \in (\pi_1({\cal N});\pi_3(M))$.
\label{table:classification-fractional}}
\end{table}
%%%%%%%%%%%%%%%%%%%%%%
We label all configurations by 
the first homotopy group of the fixed point manifold 
${\cal N}$ and the instanton (baryon) number $B$:
\beq 
 (v_1,v_2;B) \in (\pi_1({\cal N});\pi_3(M)).
\eeq
The first homotopy group $\pi_1 ({\cal M})$ 
of the moduli space does not give independent information, 
so we omit it from the label. 
The number of elementary vortices 
in the $SU(3)$ case is twice of that of the $SU(2)$ case, 
just because of the two independent  
vortices in Eq.~(\ref{eq:SU(3)-vortex1}) 
compared with one for the $SU(2)$ case.

As the case of $SU(2)$, all vortices are global vortices having 
the divergent energy
\beq
  E \sim \log \Lambda /\xi
\eeq
at large distance, apart from finite contribution 
from the core.
In the $SU(3)$ case,
 the size $\xi$ of the vortex core is 
$\xi \sim m_1^{-1},m_2^{-1},m_3^{-1}$ depending on species.

The interaction between vortices 
at large distance $\rho \gg \xi$ 
depends only on their winding numbers: 
\beq
 E_{\rm int} = \mp 2 \log \rho /\xi  ,\quad 
 F_{\rm int} = \pm {2\over \rho}  \label{eq:int1}
\eeq 
for vortices of the same kind 
$U_a(\theta)$ and $U_a(\theta)$ (the upper sign; repulsion), 
and 
for a vortex $U_a(\theta)$ and 
an anti-vortex
$U_a(-\theta)$ (the lower sign; attraction) 
of the same kind. 
It is, however, opposite for 
different kinds of vortices:  
\beq
 E_{\rm int} = \pm  \log \rho/\xi  ,\quad
 F_{\rm int} = \mp \1{\rho} ,\label{eq:int2}
\eeq 
for a vortex $U_a(\theta)$ and a vortex 
$U_b(\theta)$ ($a\neq b$) 
(the upper sign; attraction) 
and a vortex $U_a(\theta)$ and an anti-vortex 
$U_b(\theta)$ ($a\neq b$) 
(the lower sign; repulsion).
The interaction between two vortices 
at short distance $\rho \sim \xi$ 
depends on the moduli $\alpha^{\pm}$ 
in the cores, but we do not discuss it.

A unit (anti-)instanton $(0,0;\pm 1)$ can be 
${\mathbb Z}_3$ symmetrically
decomposed into 
three $\pm 1/3$ instantons in Eq.~(\ref{eq:SU(3)-vortex1}) 
with $\alpha=\alpha^+$ in Eq.~(\ref{eq:z-dep}):
\beq
&& (0,0;+1) = (+1,0;+1/3) +(0,+1;+1/3)+(-1,-1;+1/3),\non
&& (0,0;-1) = (-1,0;-1/3) +(0,+1;-2/3)+(+1,+1;-1/3).
\eeq
However, the decompositions from left to right 
are energetically unfavorable 
 at least for large radius $R$, 
because 
of the absence of vortices in the left 
and the presence of three vortices in the right.
This can be also verified from the interactions 
in Eq.~(\ref{eq:int2}) 
between two among them are attractive 
at large separations. 

A unit (anti-)instanton $(0,0;\pm 1)$ also can be decomposed asymmetrically 
into two fractions as 
\beq
&& (0,0;+1) = (+1,0;+1/3) +(-1,0;+2/3),\non
&& (0,0;+1) = (0,+1;+1/3) +(0,-1;+2/3),\non
&& (0,0;+1) = (-1,-1;+1/3) +(+1,+1;+2/3),\non
&& (0,0;-1) = (-1,0;-1/3) +(+1,0;-2/3),\non
&& (0,0;-1) = (0,-1;-1/3) +(0,+1;-2/3),\non
&& (0,0;-1) = (+1,+1;-1/3) +(-1,-1;-2/3).
\eeq
These decompositions are also energetically 
unfavorable at least for large radius $R$ of $S^1$.

The charge two (anti-)instanton $(0,0;\pm 2)$ 
also can decay 
in several ways such as
\beq
&& (0,0;+2) =  (-1,0;+2/3)+(0,-1;+2/3)+(+1,+1;+2/3),\non
&& (0,0;-2) = (+1,0;-2/3)+(0,+1;-2/3)+(-1,-1;-2/3).
\eeq
or 
\beq
 && (0,0;+2) = 2(+1,0;+1/3) +2(0,+1;+1/3)+2(-1,-1;+1/3),\non
&& (0,0;-2) = 2(-1,0;-1/3) +2(0,+1;-2/3)+2(+1,+1;-1/3).
\eeq

As one can expect, the fundamental fractional instantons 
with the instanton charge $\pm 2/3$ can be 
decomposed, at least homotopically, 
into two fractional instantons with 
the instanton charge $\pm 1/3$ as:
\beq
   (-1,0;+2/3) &=& (0,+1;+1/3) + (-1,-1;+1/3) ,\non
   (0,-1;+2/3) &=& (-1,-1;+1/3) +  (+1,0;+1/3),\non
 (+1,+1;+2/3) &=& (+1,0;+1/3) +  (0,+1;+1/3),\non
   (+1,0;-2/3) &=& (0,-1;-1/3) + (+1,+1;-1/3) ,\non
   (0,+1;-2/3) &=&  (+1,+1;-1/3) +  (-1,0;-1/3),\non
 (-1,-1;-2/3) &=&   (-1,0;-1/3) +  (0,-1;-1/3).
\eeq
These decompositions are energetically unfavorable 
because the numbers of vortices 
are one in the left and two in the right. 
Unlike the case of decomposition 
of unit instantons, 
we regard configurations with 
$B=2/3$ are elementary for the moment, 
because the vortex winding numbers are the minimum.

In order to satisfy the twisted boundary condition 
in Eq.~(\ref{eq:tbc-phase}),
one may consider a configuration with 
a more rapid $z$-dependence 
modulo $2\pi$ instead of Eq.~(\ref{eq:z-dep}),
such as $8\pi/3=2\pi+2\pi/3$. 
 However, it can be decomposed 
through a self-reconnection 
into a closed-line configuration with an integer $B$ 
that does not reach the boundary  
and a fraction given above.
In this sense, such configurations are not elementary.

A comment is in order here.
The decompositions of the unit instanton 
and $B=2/3$ instantons 
are very similar to those of vortices (color flux tubes) 
in dense QCD  
\cite{Balachandran:2005ev,Eto:2013hoa}.
The unit instanton corresponds to
a $U(1)$ superfluid vortex without color flux 
and $B=1/3$ and $2/3$ fractional instantons 
correspond to 
$M_1$ and $M_2$ non-Abelian vortices 
having color fluxes, respectively.

%%%%%%%%%%%%
\subsection{Neutral bions}\label{sec:SU(3)neutral}
As the $SU(2)$ case, 
neutral bions  are configurations with 
zero instanton charges and zero vortex charges:
\beq
 \sum_i (v_{1,i},v_{2,i};B_i) = (0,0;0) . \label{eq:neutral2}
\eeq
Let us define the order of 
neutral bions as the maximum instanton charge
of a subgroup of constituents. 

The lowest order of neutral bions is 
$1/3$ (the total instanton charge is therefore 
$B=1/3-1/3$):  
\beq
  (+1,0;+1/3) + (-1,0;-1/3) &=& (0,0;0)  ,\non
  (0,+1;+1/3) + (0,-1;-1/3) &=& (0,0;0)  ,\non
  (-1,-1;+1/3) + (+1,+1;-1/3)  &=& (0,0;0)  .
\eeq
The neutral bions 
in the left sides can annihilate in a pair to the vacuum.

There exist two kinds of 
neutral bions of the order $2/3$ 
($B=2/3-2/3$). 
The simplest ones are 
composed of two fractional 
(anti-)instantons: 
\beq
   (-1,0;+2/3) + (+1,0;-2/3) &=& (0,0;0)  ,\non
   (0,-1;+1/3) + (0,+1;-2/3) &=& (0,0;0)  ,\non
 (+1,+1;+2/3) + (-1,-1;-2/3)  &=& (0,0;0)  .
\eeq
They can annihilate in pair to the vacuum. 
More nontrivial ones are made of three 
fractional instantons:
\beq
 (+1,0;+1/3) + (0,+1;+1/3) + (-1,-1;-2/3)  &=& (0,0;0)  ,\non
 (0,+1;+1/3) + (-1,-1;+1/3) + (+1,0;-2/3)  &=& (0,0;0)  ,\non
 (-1,-1;+1/3) + (+1,0;+1/3) + (0,+1;-2/3)  &=& (0,0;0)  ,\non
  (-1,0;-1/3) + (0,-1;-1/3) + (+1,+1;-2/3)  &=& (0,0;0)  ,\non
 (0,-1;-1/3) + (+1,+1;-1/3) + (-1,0;-2/3)  &=& (0,0;0)  ,\non
 (+1,+1;-1/3) + (-1,0;+1/3) + (0,-1;-2/3)  &=& (0,0;0)  .
\eeq
Since each set of them is not a simple pair 
of fractional and anti-fractional instantons, 
it does not have to annihilate to the vacuum.
Instead it may constitute a stable bound state.

Interesting is that we have neutral bions of the order 
one ($B=1-1$) that is {\it not}  
a pair of instanton and anti-instanton:
\beq
 &&  [(+1,0;+1/3) + (0,-1;+2/3)] + [(-1,0;-1/3) + (0,+1;-2/3)] \non
&=& (+1,-1;+1) + (-1,+1;-1) = (0,0;0), \non
 && [(0,+1;+1/3) + (+1,+1;+2/3)] + [(0,-1;-1/3) + (-1,-1;-2/3)] \non
&=& (+1,+2;+1) + (-1,-2;-1)  = (0,0;0) , \non
&& [(-1,-1;+1/3) + (-1,0;+2/3)] + [(+1,+1;-1/3) + (+1,0;-2/3)] \non
&=& (-2,-1;+1) + (+2,+1;-1) = (0,0;0), \non
&&  [(+1,0;+1/3) + (+1,+1;+2/3)] +[(-1,0;-1/3) + (-1,-1;-2/3)] \non
&=& (+2,+1;+1) + (-2,-1;-1) = (0,0;0), \non
&& [(0,+1;+1/3) + (+1,+1;+2/3)] + [(0,-1;-1/3) + (-1,-1;-2/3)] \non
&=& (+1,+2;+1) + (-1,-2,-1) = (0,0;0), \non
&& [(-1,-1;+1/3) + (-1,0;+2/3)] + [(+1,+1;-1/3) + (+1,0;-2/3)]\non
&=& (-2,-1;+1) + (+2,+1;-1) = (0,0;0).
\eeq

More surprisingly, there are 
neutral bions of the order greater than 
one, that do not contain 
instanton or anti-instanton. 
For instance, the following is of the order $4/3$:
\beq
&& [ (+1,0;+1/3)+(0,+1;+1/3)+(+1,+1;+2/3) ]\non
&+& [ (-1,0;-1/3)+(0,-1;-1/3)+(-1,-1;+2/3) ]\non
&=& (+2,+2;+4/3) + (-2,-2;-4/3)
= (0,0;0).
\eeq

%%%%%%%%%%%
\subsection{Charged bions}\label{sec:SU(3)charged}
In the same way, charged bions  are configurations with 
zero instanton charges and non-zero vortex charges:
\beq
 \sum_i (v_{1,i},v_{2,i};B_i) = (v_1,v_2,0), \quad 
(v_1,v_2)\neq (0,0) . \label{eq:charged2}
\eeq
We define the order of charged bions 
as the same way with that of neutral bions.
Charged bions of the order $1/3$ are 
composed of two fractional 
(anti-)instantons 
with the topological charges $B=\pm1/3$ 
are
\beq
  (+1,0;+1/3) + (0,-1;-1/3) &=& (1,-1;0), \non
  (+1,0;+1/3) + (+1,+1;-1/3) &=& (2,+1;0) , \non
  (0,+1;+1/3) + (-1,0;-1/3) &=& (-1,+1;0) ,\non
  (0,+1;+1/3) + (+1,+1;-1/3) &=& (+1,+2;0), \non
  (-1,-1;+1/3) + (-1,0;-1/3) &=& (-2,-1;0) ,\non
  (-1,-1;+1/3) + (0,-1;-1/3) &=& (-1,-2;0) .
\eeq
We may call them ``mesons." 

Similarly to this, 
charged bions of the order $2/3$, 
which are 
composed of two fractional 
(anti-)instantons 
with the topological charges $B=\pm2/3$ are:
\beq
  (-1,0;+2/3) + (0,+1;-2/3) &=& (-1,+1;0), \non
  (-1,0;+2/3) + (-1,-1;-2/3) &=& (-2,-1;0) , \non
  (0,-1;-1/3) + (+1,0;-2/3) &=& (+1,-1;0) ,\non
  (0,-1;-1/3) + (-1,-1;-2/3) &=& (-1,-2;0), \non
  (+1,+1;-1/3) + (+1,0;-2/3) &=& (+2,+1;0) ,\non
  (+1,+1;-1/3) + (0,+1;-2/3) &=& (+1,+2;0) .
\eeq
In addition, 
there are charged bions of the order $2/3$ ($B=2/3-2/3$),
that are 
composed of three distinct fractional (anti-)instantons: 
\beq
 (+1,0;+1/3) + (0,+1;+1/3) + (+1,0;-2/3) &=& (+2,+1;0), \non
 (+1,0;+1/3) + (0,+1;+1/3) + (0,+1;-2/3) &=& (+1,+2;0), \non
 (0,+1;+1/3) + (-1,-1;+1/3) + (+1,0;-2/3) &=& (+1,+1;0), \non
 (0,+1;+1/3) + (-1,-1;+1/3) + (-1,-1;-2/3) &=& (-2,-1;0), \non
 (-1,-1;+1/3) + (+1,0;+1/3) + (-1,-1;-2/3) &=& (-1,-2;0),\non
 (-1,-1;+1/3) + (+1,0;+1/3) + (+1,0;-2/3) &=& (+1,-1;0),
\eeq
and those composed of the two same and one distinct 
(anti-)fractional instantons:
\beq
 2(+1,0;+1/3) + (+1,0;-2/3) &=& (+3,0;0),\non
 2(0,+1;+1/3) + (0,+1;-2/3) &=& (0,+3;0),\non
 2(-1,-1;+1/3) + (-1,-1;-2/3) &=& (-3,-3;0),\non
 2(+1,0;+1/3) + (0,+1;-2/3) &=& (+2,+1;0),\non
 2(+1,0;+1/3) + (-1,-1;-2/3) &=& (+1,-1;0),\non
 2(0,+1;+1/3) + (+1,0;-2/3) &=& (+1,+2;0),\non
 2(0,+1;+1/3) + (-1,-1;-2/3) &=& (-1,+1;0),\non
 2(-1,-1;+1/3) + (+1,0;-2/3) &=& (-1,-2;0),\non
 2(-1,-1;+1/3) + (0,+1;-2/3) &=& (-2,-1;0).
\eeq

%%%%%%%%%%%%%%%%%%%%%
\section{Generalization to 
the $SU(N)$ principal chiral model\label{sec:SU(N)} }

It is straightforward to generalize our results 
to the $SU(N)$ principal chiral model 
with the center symmetric twisted boundary condition in 
Eq.~(\ref{eq:tbcSU(N)}).
The fixed manifold ${\cal N}$ is 
\beq
&& U = {\rm diag.} (e^{i\alpha_1},e^{i\alpha_2},
\cdots,e^{-i\sum_{a=1}^{N-1}\alpha_a}),
\quad
{\cal N} \simeq U(1)^{N-1}. \label{eq:homotopy-SU(N)}
\eeq
The non-trivial first homotopy group 
\beq
 \pi_1 ({\cal N}) \simeq {\mathbb Z}^{N-1} 
\eeq
admits homotopically distinct $N-1$ kinds of vortices. 
The fundamental vortices in the $SU(N)$ principal chiral model can be obtained by embedding 
the one in Eq.~(\ref{eq:SU(2)-vortex2}) of 
the $SU(2)$ principal chiral model to the $SU(N)$ matrix:
\beq
&& U_1 (r,\theta,z) = 
\left(\begin{array}{ccc} 
   \cos f(r) e^{+i \theta} & - \sin f (r) e^{+i\alpha(z)} & 0\\
   \sin f(r) e^{-i\alpha(z)} &   \cos f(r) e^{-i \theta} & 0\\
   0 & 0 & {\bf 1}_{N-2}
\end{array}
\right),\non
&& U_2 (r,\theta,z) = 
\left(\begin{array}{cccc} 
1 & 0 & 0 & \\
0 & \cos f(r) e^{+i \theta} & - \sin f (r) e^{+i\alpha(z)} &\\
0 & \sin f(r) e^{-i\alpha(z)} &   \cos f(r) e^{-i \theta}  &\\
 & & & {\bf 1}_{N-3}
\end{array}
\right),\non
&& \hspace{5cm} \vdots \non
&& U_N (r,\theta,z) = 
\left(\begin{array}{ccc} 
   \cos f(r) e^{-i \theta} & 0 & + \sin f (r) e^{-i\alpha(z)} \\
 0 & {\bf 1}_{N-2} & 0\\
 -\sin f(r) e^{+i\alpha(z)} &  0 & \cos f(r) e^{+i \theta} 
\end{array}
\right).  \label{eq:SU(N)-vortex1} 
\eeq
Anti-vortices can be obtained as
$U_a(r,-\theta)$ with $a=1,\cdots,N-1$. 
These are not homotopically 
independent of each other.  
We take the first $N-1$ as 
the independent basis of the first homotopy group 
in Eq.~(\ref{eq:homotopy}), 
in which 
the topological vortex charges in 
Eq.~(\ref{eq:SU(3)-vortex1}) are ($N-1$ vectors)
\beq
&& 
 u_1=(1,0,\cdots,0), \;\;\;
 u_2=(0,1,0,\cdots,0), \quad \cdots, \non
&& u_{N-1}=(0,0,0,\cdots,0,1), \;\;\;
 u_N=(-1,-1,\cdots,-1) 
\in 
 {\mathbb Z}^{N-1}  \simeq \pi_1 ({\cal N}) ,
\eeq
respectively. 
The last one is not independent of the rests  
as can be seen from the fact that 
the vortex charges are canceled out 
when all three are present together:
\beq
 \sum_{a=1}^N u_a = (0,0,\cdots,0).
\eeq
In Eq.~(\ref{eq:SU(N)-vortex1}), 
$\alpha$ is a $U(1)$ modulus of a vortex
that is constant if the vortex does not 
wrap the $S^1$ direction.

When a vortex wraps the $S^1$ direction, 
the modulus $\alpha$ must change along the vortex 
enforced by  
the twisted boundary condition in Eq.~(\ref{eq:tbcSU(N)}):
\beq
 \exp {i\alpha (z+R) } 
=  \exp {\left[i \alpha (z) + {2\pi i\over N}\right]} ,
  \label{eq:tbc-phase-SU(N)}
\eeq
where $z$ is coordinate along $S^1$ with the period 
$R$.
This boundary condition can be satisfied by 
the two different  minimum paths with 
the following $z$-dependence of $\alpha$: 
\beq
\begin{displaystyle}
 \alpha (z) = 
\Bigg\{\begin{array}{c}
\alpha^+(z) = \alpha_0 +  \displaystyle{{2\pi\over N} {z\over R}} \\
\alpha^-(z) = \alpha_0 -  \displaystyle{{2\pi (N-1) \over N} {z\over R}}
\end{array} 
\end{displaystyle}
.  \label{eq:z-dep-SU(N)}
\eeq
Correspondingly, 
each of them carries fractional instanton (baryon) number:
\begin{align}
B 
= \pm \1{2\pi} [\alpha]^{z=R}_{z=0} =
\Bigg\{\begin{array}{c}
\pm \displaystyle{{1\over N}}   
  \quad {\rm for}\; \alpha = \alpha^+\\
\mp \displaystyle{{N-1 \over N}}  
  \quad {\rm for}\; \alpha = \alpha^-
\end{array} ,
\end{align}
where the upper and lower signs correspond to 
a vortex or anti-vortex, respectively. 
We thus have found  
$2N$ ($2N-2$ independent) types of elementary 
fractional instantons 
as well as $2N$ ($2N-2$ independent) types of 
elementary fractional anti-instantons, 
labeled by $(v_1,\cdots,v_{N-1};B)$. 

Neutral bions and charged bions can be 
constructed in the same way with 
the $SU(3)$ case, but the number of combinations 
grows drastically. 
We need a more systematic analysis 
that we leave for a future study.

%%%%%%%%%%%%%%%%%%%%%
\section{More general twisted boundary conditions}
\label{sec:general}

The more general twisted boundary condition 
for the $SU(N)$ principal chiral model was given 
in Eq.~(\ref{eq:tbcSU(N)2}).
In this case, the boundary condition on the 
$U(1)$ moduli are 
\beq
 \exp {i\alpha_a (z+R) } 
=  \exp {\left[i \alpha_a (z) 
+ {i (m_{a+1}-m_a)}\right]} ,
  \label{eq:tbc-phase-SU(N)2}
\eeq
with $m_{N+1} \equiv m_1 + 2\pi$.
This can be satisfied by the following $z$-dependence 
of the moduli:
\beq
\begin{displaystyle}
 \alpha (z) = 
\Bigg\{\begin{array}{c}
\alpha^+(z) = +  \displaystyle{(m_{a+1}-m_a) {z\over R}} \\
\alpha^-(z) = -  \displaystyle{(2\pi - m_{a+1} + m_a) {z\over R}}
\end{array} 
\end{displaystyle}
.  \label{eq:z-dep}
\eeq
Correspondingly, 
each of them carries fractional instanton (baryon) number 
that are not rational number anymore:
\begin{align}
B_a 
= \pm \1{2\pi} [\alpha_a]^{z=R}_{z=0} =
\Bigg\{\begin{array}{c}
\pm (m_{a+1}-m_a)  \quad {\rm for}\;\; \alpha = \alpha^+\\
\mp (2\pi - m_{a+1}+m_a)   \quad {\rm for}\;\; \alpha=\alpha^-
\end{array} .
\label{eq:charge-general}
\end{align}
The sum of all fractions is of course unity:
\beq
 \sum_{a=1}^N B_a = {m_{N+1} - m_1 \over 2\pi} = 1.
\eeq

In the previous sections, we have defined 
bions as configurations with zero instanton charges, 
that is enough for the ${\mathbb Z}_N$ 
symmetric twisted boundary conditions.
In Refs.~\cite{Argyres:2012ka,Dunne:2012ae}, 
an invariant definition of bions 
using the Lie algebra was given.
Here, we follow their definition of bions.
Instead of using the Lie algebra,
it is enough to regard bions in the previous sections 
remain bions for generic boundary conditions.
In this case, 
cancellation of instanton charges for bions 
becomes very restrictive 
and is drastically simplified,
since
fractions of instanton charges 
are not rational.
For non-degenerate and 
generic $m_a$ without any particular relation 
such as $m_a + m_b = m_c$, 
fractional instanton charges cannot be canceled out 
among different types of fractional instantons.
Consequently, 
only neutral bions of the minimum orders 
composed of  
a vortex and an anti-vortex of the same kind
with the same $U(1)$ moduli shift 
along the $S^1$ direction 
have zero instanton charges. 
We conclude the existence of 
$2N$ neutral bions of the minimum order, 
but higher order ones, 
having zero instanton charges for 
the ${\mathbb Z}_N$ symmetric boundary condition, 
have instanton charges.

On the other hand,
charged bions have instanton charges 
in general,
because different vortices cannot cancel 
their instanton charges unless a particular 
relation such as $m_a + m_b =m_c$ exists.

A partially degenerated case
is interesting 
since vortices would carry non-Abelian moduli.
We will return to this case in a future.

%%%%%%%%%%%%%%%%%%%%%
\section{Relation to 
Yang-Mills fractional instantons and bions\label{sec:YM} }

A ${\mathbb C}P^{N-1}$ instanton 
with the twisted boundary condition 
is decomposed into a set of $N$ fractional instantons 
which are half twisted domain walls.
The same relation holds between 
a Yang-Mills instanton and a BPS monopole. 
In Ref.~\cite{Eto:2004rz,Eto:2006pg,Fujimori:2008ee}, 
 ${\mathbb C}P^{N-1}$ fractional instantons 
were realized as  fractional Yang-Mills instantons 
trapped inside a vortex in a $U(N)$ gauge theory.
This explains a correspondence between 
fractional instantons and bions 
in the ${\mathbb C}P^{N-1}$ 
on ${\mathbb R}^1 \times S^1$ 
and those in $SU(N)$ Yang-Mills in 
on ${\mathbb R}^3 \times S^1$. 
In this section, we point out 
 fractional instantons and 
bions in the $SU(N)$ principal chiral model 
also correspond to 
$SU(N)$ Yang-Mills 
fractional instantons and bions.

Let us consider the $U(N)$ gauge theory 
 in $d=4+1$ dimensions  ($A,B=0,\cdots ,4$). 
The matter contents are 
$U(N)$ gauge field $A_A(x)$, 
$2N$ complex scalar fields 
$H (x) = (H_1(x),H_2(x))$  (decomposed into 
two $N\times N$ matrices of scalar fields) 
in the fundamental representation 
charged under $U(1)$,
an $N \times N$ matrix  $\Sigma(x)$  
of scalar fields
in the adjoint representation 
neutral under $U(1)$. 
The Lagrangian is given by  
\cite{Shifman:2003uh,Eto:2005cc,Eto:2008dm} 
\beq
&& {\cal L} \,=\, -\1{4 g^2} \tr F_{AB} F^{AB} 
 + \1{g^2} \tr (D_A\Sigma)^2 \
 + \tr |D_A H|^2  - V  \,\\
&& V \,=\, {g^2 \over 4} \tr ( H H^\dagger 
 \ - \ v^2 {\bf 1}_N)^2 
 + \tr |\Sigma H - H M|^2 ,\,
\eeq
where the covariant derivatives are given by  
$D_A H = \del_A H - i A_A H$ 
and 
$D_A \Sigma = \del_A \Sigma - i [A_A,\Sigma]$, 
$g$ is the gauge coupling constant that we take 
common for 
the $U(1)$ and $SU(N)$ factors of $U(N)$, 
$v$ is a real constant giving the vacuum expectation value 
of $H$,  
$M$ is a $2N$ by $2N$ mass matrix given as
 $M={\rm diag.}(m {\bf 1}_N ,
- m {\bf 1}_N 
)$.
The $U(N)$ gauge (color) symmetry and 
the flavor symmetry act on fields as 
\beq
&& A_A \to g A_A g^{-1} + i g \del_A g^{-1},  \quad
 \Sigma \to g \Sigma g^{-1}, \quad
 H \to g H \quad g \in U(N)_{\rm C}, \\
 && 
H_1 \to H_1 U_{\rm L} e^{+i \alpha}, \quad 
 H_2 \to H_2 U_{\rm R} e^{-i \alpha},  \quad U_{\rm L,R} \in SU(N)_{\rm L,R}.
\label{eq:flavor}
\eeq
The model admits two disjoint vacua 
\beq
&& H \ = \
\left(
 v {\bf 1}_N ,{\bf 0}_N 
\right) , \quad
\Sigma = - m {\bf 1}_N:  
\ \quad \ SU(N)_{\rm C+L} ,
\non
&& H \ = \  
\left(
  {\bf 0}_N , v{\bf 1}_N 
\right) , 
\  \quad \ \Sigma = + m {\bf 1}_N:
\quad \ SU(N)_{\rm C+R} 
 \label{eq:vac}
\eeq 
with the unbroken color-flavor locked {\it global} symmetries
$g=U_{\rm L}$ and $g = U_{\rm R}$, respectively.

The model admits a non-Abelian 
domain wall solution
interpolating between the two vacua 
in Eq.~(\ref{eq:vac}),  
that is obtained by 
embedding the ${\mathbb C}P^1$ domain wall  \cite{Abraham:1992vb}.
The solution perpendicular to the $x^4$ coordinate  
can be given by 
 \cite{Isozumi:2004jc,Shifman:2003uh,Eto:2005cc,Eto:2008dm}
\beq 
&& H_{\rm wall} \ = \ V H_{\rm wall,0} 
\left(\begin{array}{cc}
 V^\dagger & 0 \\ 0 & V
\end{array} \right) 
=  {v\over \sqrt {1+ e^{\mp 2 m (x^4-X) }}}
      \left({\bf 1}_N, e^{\mp m (x^4-X) }U \right), 
\non
&&
 \Sigma_{\rm wall} \ = \ V \Sigma_{\rm wall,0} V^\dagger ,
\quad
A_{4,{\rm wall}} \ = \ V A_{4,{\rm wall},0} V^\dagger ,
\eeq 
with $V \in SU(N)$, 
and we have defined the moduli  
$U \equiv V^2 e^{i\ph} \in U(N)$ of the domain wall:
\beq 
 {\cal M}_{\rm wall} \simeq {\mathbb R} \times U(N) 
\ni (X,U).
\label{eq:wall-moduli}
\eeq 
The width of the domain wall is $m^{-1}$.

The low-energy dynamics of 
of the non-Abelian 
domain wall
can be described by 
the effective theory 
within the moduli approximation \cite{Manton:1981mp,Eto:2006uw}, 
by promoting the moduli parameters 
$X$ and $U$ to moduli fields 
 $X (x^\mu)$ and $U(x^\mu)$, respectively 
($\mu=0,1,2,3$) 
on the world volume of the domain wall,  
and by performing integration over the codimension.  
We thus obtain the effective theory  \cite{Shifman:2003uh,Eto:2005cc,Eto:2008dm}:
\beq
 && {\cal L}_{\rm wall} =  
{v^2 \over 2m} \del_\mu X \del^i X 
-  f_{\pi}^2 
\tr \left(U^\dagger \del_\mu U
            U^\dagger \del^\mu U \right) ,
\quad  f_{\pi}^2 \equiv {v^2 \over 4m} 
\label{eq:eff},
\eeq
that is a $U(N)$ principal chiral model we are discussing. 

It was shown in Ref.~\cite{Eto:2005cc} that 
Yang-Mills instantons inside the domain wall 
are described by 
Skyrmions in the principal chiral model 
on the domain wall.  
This setting physically realizes the 
Atiyah-Manton construction of Skyrmions 
from instanton holonomy \cite{Atiyah:1989dq}. 
Furthermore, it has been recently found 
in Ref.~\cite{Nitta:2014vpa} that magnetic monopoles 
inside the domain wall are described by 
vortices in the principal chiral model 
with the twisted mass term in Eq.~(\ref{eq:twisted-mass}). 
In fact, the instanton charge $\pi_3$ in Eq.~(\ref{eq:charge-general}) 
is proportional to the monopole charge. 
Interpolating solutions for the $SU(2)$ case 
were already constructed numerically in the Skyrme model 
with the twisted boundary condition \cite{Harland:2008eu}.
In the limit of $m \to 0$, the width of the domain wall 
diverges and further taking $v \to 0$, 
the system leaves from the Higgs phase recovering 
the pure Yang-Mills theory.  
The fractional instantons and bions become 
those in Yang-Mills theory.
Therefore, this setting offers 
a correspondence between 
fractional instantons and bions 
in the $SU(N)$ principal chiral model 
and those in Yang-Mills theory. 

%%%%%%%%%%%%%%%%%%%%%
\section{Summary and Discussion \label{sec:summary} }

We have classified fractional instantons and bions 
in the $SU(N)$ principal chiral model 
on ${\mathbb R}^2 \times S^1$ with 
twisted boundary conditions.
This model allows 
$N$ kinds of global vortices 
with $U(1)$ moduli, 
among which $N-1$ kinds are independent. 
Fractional instantons 
are $N$  ($N-1$ independent) kinds of global vortices wrapping around $S^1$
with $U(1)$ moduli twisted with 
the angle $2\pi/N$ along the world-line $S^1$. 
We have found that 
they carry $1/N$ instanton (baryon) numbers
for the ${\mathbb Z}_N$ symmetric twisted boundary condition 
and irrational instanton numbers 
for the generic  boundary conditions 
with the general phases $m_a$.  
We have classified neutral and charged bions 
for the $SU(3)$ case with 
the ${\mathbb Z}_3$ symmetric twisted boundary condition.
We have found for the generic boundary conditions 
that there are
only the simplest neutral bions 
composed of a pair of (anti-)fractional instantons 
have zero instanton charges   
and instanton charges are not canceled out for 
charged bions 
unless some particular relation holds among $m_a$. 
We have also discussed 
a correspondence between 
fractional instantons and bions 
in the $SU(N)$ principal chiral model 
and those in Yang-Mills theory 
through a non-Abelian Josephson junction.

We have studied ${\mathbb Z}_N$ center 
symmetric twisted boundary condition and 
non-degenerate ($m_a \neq m_b$ for $a \neq b$) 
boundary conditions.
A partially degenerated case
will be interesting 
since vortices would carry non-Abelian moduli 
in this case. 
We will return to this case in a future.

In this paper, we have put the twisted boundary conditions 
by hand, but 
the ${\mathbb Z}_N$ symmetric boundary condition 
was chosen  for 
the ${\mathbb C}P^{N-1}$ model 
from the effective potential in quantum theory
\cite{Dunne:2012ae,Dunne:2012zk}.
The same analysis should be 
done in our case. Although the principal chiral 
model is not renormalizable in perturbation  
in three dimensions, 
they are renormalizable at large $N$. 
We will return to this problem in the future.

In order to generalize to the principal chiral model with 
arbitrary groups, 
the description of bions in terms of the Lie algebra 
and the root system given in 
Refs.~\cite{Argyres:2012ka,Dunne:2012ae} 
should be useful.

In the context of the Skyrme model,
Skyrmion chains on ${\mathbb R}\times S^1$ 
with a twisted boundary condition were studied 
before \cite{Harland:2008eu},
in which a vortex structure was found. 
If we add the Skyrme term to our model, 
we can consider 
$SU(N)$ Skyrmion chains.

The fractional instantons in the principal chiral model
are all global vortices and 
the interaction between them is long range, 
$E_{\rm int} \sim \pm \log \rho$ with the distance $\rho$.
Therefore, they are confined.
If we gauge the $U(1)^{N-1}$ center action,
vortices become local vortices, {\it i.~e.~}, 
of the Abrikosov-Nielsen-Olesen type \cite{Abrikosov:1956sx}, 
for which 
the interaction is exponentially suppressed 
with respect to the distance. 
While this case will be interesting on its own, 
fractional instantons are also local 
and the total action is the sum of 
actions of individual fractional instantons so that 
they would be useful for resurgence of the quantum 
field theory. 

Supersymmetric extension is possible 
by generalizing the target space to 
the cotangent bundle, $T^*SU(N)$, that is K\"ahler.

%%%%%%%%%%%%%%
\section*{Acknowledgments}

The author thanks Tatsuhiro Misumi and Norisuke Sakai 
for a discussion on bions. 
This work is supported in part by Grant-in-Aid for Scientific Research 
No.~25400268
and by the ``Topological Quantum Phenomena'' 
Grant-in-Aid for Scientific Research 
on Innovative Areas (No.~25103720)  
from the Ministry of Education, Culture, Sports, Science and Technology 
(MEXT) of Japan.

%%%%%%%%%% References %%%%%%%%%%%%%%%%%%%%%%%%%
\newcommand{\J}[4]{{\sl #1} {\bf #2} (#3) #4}
\newcommand{\andJ}[3]{{\bf #1} (#2) #3}
\newcommand{\AP}{Ann.\ Phys.\ (N.Y.)}
\newcommand{\MPL}{Mod.\ Phys.\ Lett.}
\newcommand{\NP}{Nucl.\ Phys.}
\newcommand{\PL}{Phys.\ Lett.}
\newcommand{\PR}{ Phys.\ Rev.}
\newcommand{\PRL}{Phys.\ Rev.\ Lett.}
\newcommand{\PTP}{Prog.\ Theor.\ Phys.}
\newcommand{\hep}[1]{{\tt hep-th/{#1}}}
%%%%%%%%%%%%%%%%%%%%%%%%%%%%%%%%%%%%%%%%%%%%%%%


\begin{thebibliography}{100}




\bibitem{Unsal:2007vu} 
  M.~\"{U}nsal,
  ``Abelian duality, confinement, and chiral symmetry breaking in QCD(adj),''
  Phys.\ Rev.\ Lett.\  {\bf 100}, 032005 (2008)
  [arXiv:0708.1772 [hep-th]].
  %%CITATION = ARXIV:0708.1772;%%
  
\bibitem{Unsal:2007jx} 
  M.~\"{U}nsal,
  ``Magnetic bion condensation: A New mechanism of confinement and mass gap in four dimensions,''
  Phys.\ Rev.\ D {\bf 80}, 065001 (2009)
  [arXiv:0709.3269 [hep-th]].
  %%CITATION = ARXIV:0709.3269;%%

\bibitem{Shifman:2008ja} 
  M.~Shifman and M.~\"{U}nsal,
  ``QCD-like Theories on R(3) x S(1): A Smooth Journey from Small to Large r(S(1)) with Double-Trace Deformations,''
  Phys.\ Rev.\ D {\bf 78}, 065004 (2008)
  [arXiv:0802.1232 [hep-th]].
  %%CITATION = ARXIV:0802.1232;%%


\bibitem{Poppitz:2009uq} 
  E.~Poppitz and M.~\"{U}nsal,
  ``Conformality or confinement: (IR)relevance of topological excitations,''
  JHEP {\bf 0909}, 050 (2009)
  [arXiv:0906.5156 [hep-th]].
  %%CITATION = ARXIV:0906.5156;%%

%\cite{Anber:2011de}
\bibitem{Anber:2011de} 
  M.~M.~Anber and E.~Poppitz,
  ``Microscopic Structure of Magnetic Bions,''
  JHEP {\bf 1106}, 136 (2011)
  [arXiv:1105.0940 [hep-th]].
  %%CITATION = ARXIV:1105.0940;%%
  %14 citations counted in INSPIRE as of 09 Sep 2014


\bibitem{Poppitz:2012sw} 
  E.~Poppitz, T.~Schaefer and M.~\"{U}nsal,
  ``Continuity, Deconfinement, and (Super) Yang-Mills Theory,''
  JHEP {\bf 1210}, 115 (2012)
  [arXiv:1205.0290 [hep-th]].
  %%CITATION = ARXIV:1205.0290;%%

\bibitem{Argyres:2012vv} 
  P.~Argyres and M.~\"{U}nsal,
  ``A semiclassical realization of infrared renormalons,''
  Phys.\ Rev.\ Lett.\  {\bf 109}, 121601 (2012)
  [arXiv:1204.1661 [hep-th]].
  %%CITATION = ARXIV:1204.1661;%%
  
  \bibitem{Argyres:2012ka} 
  P.~C.~Argyres and M.~\"{U}nsal,
  ``The semi-classical expansion and resurgence in gauge theories: new perturbative, instanton, bion, and renormalon effects,''
  JHEP {\bf 1208}, 063 (2012)
  [arXiv:1206.1890 [hep-th]].
  %%CITATION = ARXIV:1206.1890;%%

\bibitem{Dunne:2012ae} 
  G.~V.~Dunne and M.~\"{U}nsal,
  ``Resurgence and Trans-series in Quantum Field Theory: The CP(N-1) Model,''
  JHEP {\bf 1211}, 170 (2012)
  [arXiv:1210.2423 [hep-th]].
  %%CITATION = ARXIV:1210.2423;%%

\bibitem{Dunne:2012zk} 
  G.~V.~Dunne and M.~\"{U}nsal,
  ``Continuity and Resurgence: towards a continuum definition of the CP(N-1) model,''
  Phys.\ Rev.\ D {\bf 87}, 025015 (2013)
  [arXiv:1210.3646 [hep-th]].
  %%CITATION = ARXIV:1210.3646;%%
  
\bibitem{Dabrowski:2013kba} 
  R.~Dabrowski and G.~V.~Dunne,
  ``Fractionalized Non-Self-Dual Solutions in the CP(N-1) Model,''
  Phys.\ Rev.\ D {\bf 88}, 025020 (2013)
  [arXiv:1306.0921 [hep-th]].
  %%CITATION = ARXIV:1306.0921;%%
  
\bibitem{Dunne:2013ada} 
  G.~V.~Dunne and M.~\"{U}nsal,
  ``Generating Non-perturbative Physics from Perturbation Theory,''
  Phys.\ Rev.\ D {\bf 89}, 041701 (2014)
  [arXiv:1306.4405 [hep-th]].
  %%CITATION = ARXIV:1306.4405;%%
  
\bibitem{Cherman:2013yfa} 
  A.~Cherman, D.~Dorigoni, G.~V.~Dunne and M.~\"{U}nsal,
  ``Resurgence in QFT: Unitons, Fractons and Renormalons in the Principal Chiral Model,''
  Phys.\ Rev.\ Lett.\  {\bf 112}, 021601 (2014)
  [arXiv:1308.0127 [hep-th]].
  %%CITATION = ARXIV:1308.0127;%%
  
\bibitem{Basar:2013eka} 
  G.~Basar, G.~V.~Dunne and M.~\"{U}nsal,
  ``Resurgence theory, ghost-instantons, and analytic continuation of path integrals,''
  JHEP {\bf 1310}, 041 (2013)
  [arXiv:1308.1108 [hep-th]].
  %%CITATION = ARXIV:1308.1108;%%

\bibitem{Dunne:2014bca} 
  G.~V.~Dunne and M.~\"{U}nsal,
  ``Uniform WKB, Multi-instantons, and Resurgent Trans-Series,''
  Phys.\ Rev.\ D {\bf 89}, 105009 (2014)
  [arXiv:1401.5202 [hep-th]].
  %%CITATION = ARXIV:1401.5202;%%

\bibitem{Cherman:2014ofa} 
  A.~Cherman, D.~Dorigoni and M.~\"{U}nsal,
  ``Decoding perturbation theory using resurgence: Stokes phenomena, new saddle points and Lefschetz thimbles,''
  arXiv:1403.1277 [hep-th].
  %%CITATION = ARXIV:1403.1277;%%

%\cite{Bolognesi:2013tya}
\bibitem{Bolognesi:2013tya} 
  S.~Bolognesi and W.~Zakrzewski,
  ``Clustering and decomposition for non-BPS solutions of the $\mathbb{CP}^{N-1}$ models,''
  Phys.\ Rev.\ D {\bf 89}, no. 6, 065013 (2014)
  [arXiv:1310.8247 [hep-th]].
  %%CITATION = ARXIV:1310.8247;%%
  %3 citations counted in INSPIRE as of 18 Nov 2014

%\cite{Misumi:2014jua}
\bibitem{Misumi:2014jua} 
  T.~Misumi, M.~Nitta and N.~Sakai,
  ``Neutral bions in the ${\mathbb C}P^{N-1}$ model,''
  JHEP {\bf 1406}, 164 (2014)
  [arXiv:1404.7225 [hep-th]];
  %%CITATION = ARXIV:1404.7225;%%
  %3 citations counted in INSPIRE as of 18 Nov 2014
%\cite{Misumi:2014rsa}
%\bibitem{Misumi:2014rsa} 
  T.~Misumi, M.~Nitta and N.~Sakai,
  ``Neutral bions in the ${\mathbb C}P^{N-1}$ model for resurgence,''
  arXiv:1412.0861 [hep-th].
  %%CITATION = ARXIV:1412.0861;%%

%\cite{Misumi:2014bsa}
\bibitem{Misumi:2014bsa} 
  T.~Misumi, M.~Nitta and N.~Sakai,
  ``Classifying bions in Grassmann sigma models and non-Abelian gauge theories by D-branes,''
  PTEP {\bf 2015}, no. 3, 033B02
  [arXiv:1409.3444 [hep-th]].
  %%CITATION = ARXIV:1409.3444;%%
  %7 citations counted in INSPIRE as of 15 Mar 2015

%\cite{Dunne:2015ywa}
\bibitem{Dunne:2015ywa} 
  G.~V.~Dunne and M.~Unsal,
  ``Resurgence and Dynamics of O(N) and Grassmannian Sigma Models,''
  arXiv:1505.07803 [hep-th].
  %%CITATION = ARXIV:1505.07803;%%

 \bibitem{Misumi:2014raa} 
  T.~Misumi and T.~Kanazawa,
  ``Adjoint QCD on $\mathbb{R}^3\times S^1$ with twisted fermionic boundary conditions,''
  JHEP {\bf 1406}, 181 (2014)
  [arXiv:1405.3113 [hep-ph]].
  %%CITATION = ARXIV:1405.3113;%% 

%\cite{Anber:2014sda}
\bibitem{Anber:2014sda} 
  M.~M.~Anber and T.~Sulejmanpasic,
  ``The absence of IR renormalons in gauge theories on $\mathbb R^3\times \mathbb S^1$ and what it means for resurgence,''
  arXiv:1410.0121 [hep-th].
  %%CITATION = ARXIV:1410.0121;%%

%\cite{Shermer:2014wxa}
\bibitem{Shermer:2014wxa} 
  S.~Shermer,
  ``Twisted CP(N-1) instanton projectors and the N-level quantum density matrix,''
  arXiv:1412.3185 [hep-th].
  %%CITATION = ARXIV:1412.3185;%%

%\cite{'tHooft:1977am}
\bibitem{'tHooft:1977am} 
  G.~'t Hooft,
  ``Can We Make Sense Out of Quantum Chromodynamics?,''
  Subnucl.\ Ser.\  {\bf 15}, 943 (1979).
  %%CITATION = SUSEE,15,943;%%
  %14 citations counted in INSPIRE as of 10 Sep 2014


\bibitem{Fateev:1994ai} 
  V.~A.~Fateev, V.~A.~Kazakov and P.~B.~Wiegmann,
  ``Principal chiral field at large N,''
  Nucl.\ Phys.\ B {\bf 424}, 505 (1994)
  [hep-th/9403099].
  %%CITATION = HEP-TH/9403099;%%
  
\bibitem{Fateev:1994dp} 
  V.~A.~Fateev, P.~B.~Wiegmann and V.~A.~Kazakov,
  ``Large N chiral field in two-dimensions,''
  Phys.\ Rev.\ Lett.\  {\bf 73}, 1750 (1994).
  %%CITATION = PRLTA,73,1750;%%
    
\bibitem{Ec1}
J.~Ecalle, ``Les Fonctions Resurgentes", Vol.~I - III 
(Publ. Math. Orsay, 1981).

%\cite{Eto:2004rz}
\bibitem{Eto:2004rz} 
  M.~Eto, Y.~Isozumi, M.~Nitta, K.~Ohashi and N.~Sakai,
  ``Instantons in the Higgs phase,''
  Phys.\ Rev.\ D {\bf 72}, 025011 (2005)
  [hep-th/0412048].
  %%CITATION = HEP-TH/0412048;%%
  %80 citations counted in INSPIRE as of 04 Apr 2014

%\cite{Eto:2006pg}
\bibitem{Eto:2006pg} 
  M.~Eto, Y.~Isozumi, M.~Nitta, K.~Ohashi and N.~Sakai,
  ``Solitons in the Higgs phase: The Moduli matrix approach,''
  J.\ Phys.\ A {\bf 39}, R315 (2006)
  [hep-th/0602170].
  %%CITATION = HEP-TH/0602170;%%
  %216 citations counted in INSPIRE as of 19 mar 2015


%\cite{Bruckmann:2007zh}
\bibitem{Bruckmann:2007zh} 
  F.~Bruckmann,
  ``Instanton constituents in the O(3) model at finite temperature,''
  Phys.\ Rev.\ Lett.\  {\bf 100}, 051602 (2008)
  [arXiv:0707.0775 [hep-th]]; 
  %%CITATION = ARXIV:0707.0775;%%
  %18 citations counted in INSPIRE as of 04 Apr 2014
%\cite{Brendel:2009mp}
%\bibitem{Brendel:2009mp} 
  W.~Brendel, F.~Bruckmann, L.~Janssen, A.~Wipf and C.~Wozar,
  ``Instanton constituents and fermionic zero modes in twisted CP**n models,''
  Phys.\ Lett.\ B {\bf 676}, 116 (2009)
  [arXiv:0902.2328 [hep-th]];
  %%CITATION = ARXIV:0902.2328;%%
  %12 citations counted in INSPIRE as of 04 Apr 2014
%\cite{Harland:2009mf}
%\bibitem{Harland:2009mf} 
  D.~Harland,
  ``Kinks, chains, and loop groups in the CP**n sigma models,''
  J.\ Math.\ Phys.\  {\bf 50}, 122902 (2009)
  [arXiv:0902.2303 [hep-th]];
  %%CITATION = ARXIV:0902.2303;%%
  %9 citations counted in INSPIRE as of 04 Apr 2014
%\cite{Bruckmann:2014sla}
%\bibitem{Bruckmann:2014sla} 
  F.~Bruckmann and T.~Sulejmanpasic,
  ``Nonlinear sigma models at nonzero chemical potential: breaking up instantons and the phase diagram,''
  Phys.\ Rev.\ D {\bf 90}, no. 10, 105010 (2014)
  [arXiv:1408.2229 [hep-th]].
  %%CITATION = ARXIV:1408.2229;%%
  %1 citations counted in INSPIRE as of 14 Dec 2014


%\cite{Eto:2006mz}
\bibitem{Eto:2006mz} 
  M.~Eto, T.~Fujimori, Y.~Isozumi, M.~Nitta, K.~Ohashi, K.~Ohta and N.~Sakai,
  ``Non-Abelian vortices on cylinder: Duality between vortices and walls,''
  Phys.\ Rev.\ D {\bf 73}, 085008 (2006)
  [hep-th/0601181]; 
  %%CITATION = HEP-TH/0601181;%%
  %47 citations counted in INSPIRE as of 04 Apr 2014
%\cite{Eto:2007aw}
%\bibitem{Eto:2007aw} 
  M.~Eto, T.~Fujimori, M.~Nitta, K.~Ohashi, K.~Ohta and N.~Sakai,
  ``Statistical mechanics of vortices from D-branes and T-duality,''
  Nucl.\ Phys.\ B {\bf 788}, 120 (2008)
  [hep-th/0703197].
  %%CITATION = HEP-TH/0703197;%%
  %28 citations counted in INSPIRE as of 28 Apr 2014


%\cite{Nitta:2014vpa}
\bibitem{Nitta:2014vpa} 
  M.~Nitta,
  ``Fractional instantons and bions in the O(N) model with twisted boundary conditions,''
  JHEP (to appear) [arXiv:1412.7681 [hep-th]].
  %%CITATION = ARXIV:1412.7681;%%
  %3 citations counted in INSPIRE as of 10 Mar 2015


\bibitem{Skyrme:1962vh} 
  T.~H.~R.~Skyrme,
  ``A Unified Field Theory of Mesons and Baryons,''
  Nucl.\ Phys.\  {\bf 31}, 556 (1962);  
  %%CITATION = NUPHA,31,556;%%
  %\bibitem{Skyrme:1961vq} 
  %T.~H.~R.~Skyrme,
  ``A Nonlinear field theory,''  
  Proc.\ Roy.\ Soc.\ Lond.\ A {\bf 260}, 127 (1961).  
  %%CITATION = PRSLA,A260,127;%%

%\cite{Shifman:2003uh}
\bibitem{Shifman:2003uh} 
  M.~Shifman and A.~Yung,
  ``Localization of nonAbelian gauge fields on domain walls at weak coupling (D-brane prototypes II),''  Phys.\ Rev.\ D {\bf 70}, 025013 (2004)  [hep-th/0312257].  %%CITATION = HEP-TH/0312257;%%

%\cite{Eto:2005cc}
\bibitem{Eto:2005cc} 
  M.~Eto, M.~Nitta, K.~Ohashi and D.~Tong,
  ``Skyrmions from instantons inside domain walls,''  Phys.\ Rev.\ Lett.\  {\bf 95}, 252003 (2005)  [hep-th/0508130].  %%CITATION = HEP-TH/0508130;%%

%\cite{Eto:2008dm}
\bibitem{Eto:2008dm} 
  M.~Eto, T.~Fujimori, M.~Nitta, K.~Ohashi and N.~Sakai,
  ``Domain walls with non-Abelian clouds,''  Phys.\ Rev.\ D {\bf 77}, 125008 (2008)  [arXiv:0802.3135 [hep-th]].  %%CITATION = ARXIV:0802.3135;%%

%\cite{Nitta:2015mma}
\bibitem{Nitta:2015mma} 
  M.~Nitta,
  ``Josephson junction of non-Abelian superconductors and non-Abelian Josephson vortices,''
  Nucl.\ Phys.\ B {\bf 899}, 78 (2015)
  [arXiv:1502.02525 [hep-th]].
  %%CITATION = ARXIV:1502.02525;%%
  %2 citations counted in INSPIRE as of 19 Aug 2015

%\cite{Nitta:2015mxa}
\bibitem{Nitta:2015mxa} 
  M.~Nitta,
  ``Josephson instantons and Josephson monopoles in a non-Abelian Josephson junction,''
  Phys.\ Rev.\ D {\bf 92}, no. 4, 045010 (2015)
  [arXiv:1503.02060 [hep-th]].
  %%CITATION = ARXIV:1503.02060;%%
  %2 citations counted in INSPIRE as of 19 Aug 2015

%%%%%%%%%%%%%%%%%%%%%%%%%% intro up to here

\bibitem{Gudnason:2014hsa} 
  S.~B.~Gudnason and M.~Nitta,
  ``Incarnations of Skyrmions,''
  Phys.\ Rev.\ D {\bf 90}, 085007 (2014)
  [arXiv:1407.7210 [hep-th]].
  %%CITATION = ARXIV:1407.7210;%%

\bibitem{Kobayashi:2013aza} 
  M.~Kobayashi and M.~Nitta,
  ``Winding Hopfions on R$^{2} \times S^{1}$,''
  Nucl.\ Phys.\ B {\bf 876}, 605 (2013)
  [arXiv:1305.7417 [hep-th]].
  %%CITATION = ARXIV:1305.7417;%%
  %2 citations counted in INSPIRE as of 30 Oct 2014



\bibitem{Davis:1988jq} 
  R.~L.~Davis and E.~P.~S.~Shellard,
  ``The Physics Of Vortex Superconductivity. 2,''  
  Phys.\ Lett.\ B {\bf 209}, 485 (1988);  
  %%CITATION = PHLTA,B209,485;%%  
  %\bibitem{Davis:1988ij} 
  %R.~L.~Davis and E.~P.~S.~Shellard,
  ``Cosmic Vortons,''  
  Nucl.\ Phys.\ B {\bf 323}, 209 (1989);
  %%CITATION = NUPHA,B323,209;%%  
%\bibitem{Vilenkin:2000}
  A.~Vilenkin and E.~P.~S.~Shellard, 
  {\it Cosmic Strings and Other Topological Defects}, 
  (Cambridge Monographs on Mathematical Physics), 
  Cambridge University Press (July 31, 2000);
%\bibitem{Radu:2008pp} 
  E.~Radu and M.~S.~Volkov,
  ``Existence of stationary, non-radiating ring solitons in field
  theory: knots and vortons,''  
  Phys.\ Rept.\  {\bf 468}, 101 (2008)  [arXiv:0804.1357 [hep-th]];  
  %%CITATION = ARXIV:0804.1357;%%
%\bibitem{Garaud:2013iba} 
  J.~Garaud, E.~Radu and M.~S.~Volkov,
  ``Stable Cosmic Vortons,''
  Phys.\ Rev.\ Lett.\  {\bf 111}, 171602 (2013)
  [arXiv:1303.3044 [hep-th]].
  %%CITATION = ARXIV:1303.3044;%%

\bibitem{Ruostekoski:2001fc} 
  J.~Ruostekoski and J.~R.~Anglin,
  ``Creating vortex rings and three-dimensional skyrmions in
  Bose-Einstein condensates,'' 
  Phys.\ Rev.\ Lett.\  {\bf 86}, 3934 (2001)
  [cond-mat/0103310];  
  %%CITATION = COND-MAT/0103310;%%
  %\bibitem{Battye:2001ec} 
  R.~A.~Battye, N.~R.~Cooper and P.~M.~Sutcliffe,
  ``Stable skyrmions in two-component Bose-Einstein condensates,''  
  Phys.\ Rev.\ Lett.\  {\bf 88}, 080401 (2002)
  [cond-mat/0109448]; 
  %%CITATION = COND-MAT/0109448;%%
 %\bibitem{Savage:2003hh} 
  C.~M.~Savage and J.~Ruostekoski,
  ``Energetically stable particle-like skyrmions in a trapped
  Bose-Einstein condensate,''
  Phys.\ Rev.\ Lett.\  {\bf 91}, 010403 (2003);
  [cond-mat/0306112];  
  %%CITATION = COND-MAT/0306112;%%
  %\bibitem{Ruostekoski:2004pj} 
  J.~Ruostekoski,
  ``Stable particlelike solitons with multiply-quantized vortex lines
  in Bose-Einstein condensates,'' 
  Phys.\ Rev.\ A {\bf 70}, 041601 (2004);
  [cond-mat/0408376].  
  %%CITATION = COND-MAT/0408376;%%
  %\bibitem{Wuster:2005}
  S.~Wuster, T.~E.~Argue, and C.~M.~Savage,
  ``Numerical study of the stability of skyrmions in Bose-Einstein
  condensates,'' 
  Phys.\ Rev.\ A {\bf 72}, 043616 (2005);
%\bibitem{Kawakami:2012zw} 
  T.~Kawakami, T.~Mizushima, M.~Nitta and K.~Machida,
  ``Stable Skyrmions in SU(2) Gauged Bose-Einstein Condensates,''  
  Phys.\ Rev.\ Lett.\  {\bf 109}, 015301 (2012)  
  [arXiv:1204.3177 [cond-mat.quant-gas]].  
  %%CITATION = ARXIV:1204.3177;%%

\bibitem{Nitta:2012hy} 
  M.~Nitta, K.~Kasamatsu, M.~Tsubota and H.~Takeuchi,
  ``Creating vortons and three-dimensional skyrmions from domain wall
  annihilation with stretched vortices in Bose-Einstein condensates,''
  Phys.\ Rev.\ A {\bf 85}, 053639 (2012)  
  [arXiv:1203.4896 [cond-mat.quant-gas]].  
  %%CITATION = ARXIV:1203.4896;%% 

\bibitem{Metlitski:2003gj} 
  M.~A.~Metlitski and A.~R.~Zhitnitsky,
  ``Vortex rings in two-component Bose-Einstein condensates,''  
  JHEP {\bf 0406}, 017 (2004)
  [arXiv:cond-mat/0307559].  
  %%CITATION = COND-MAT/0307559;%%



\bibitem{Gudnason:2014gla} 
  S.~B.~Gudnason and M.~Nitta,
  ``Effective field theories on solitons of generic shapes,''
  arXiv:1407.2822 [hep-th].
  %%CITATION = ARXIV:1407.2822;%%
  %1 citations counted in INSPIRE as of 23 Oct 2014

%\cite{Gudnason:2014jga}
\bibitem{Gudnason:2014jga} 
  S.~B.~Gudnason and M.~Nitta,
  ``Baryonic torii: Toroidal baryons in a generalized Skyrme model,''
  Phys.\ Rev.\ D {\bf 91}, no. 4, 045027 (2015)
  [arXiv:1410.8407 [hep-th]].
  %%CITATION = ARXIV:1410.8407;%%
  %6 citations counted in INSPIRE as of 15 mar 2015

%\cite{Eto:2006db}
\bibitem{Eto:2006db} 
  M.~Eto, K.~Hashimoto, G.~Marmorini, M.~Nitta, K.~Ohashi and W.~Vinci,
  ``Universal Reconnection of Non-Abelian Cosmic Strings,''
  Phys.\ Rev.\ Lett.\  {\bf 98}, 091602 (2007)
  [hep-th/0609214].
  %%CITATION = HEP-TH/0609214;%%
  %78 citations counted in INSPIRE as of 17 Dec 2014

%\cite{Balachandran:2005ev}
\bibitem{Balachandran:2005ev} 
  A.~P.~Balachandran, S.~Digal and T.~Matsuura,
  ``Semi-superfluid strings in high density QCD,''
  Phys.\ Rev.\ D {\bf 73}, 074009 (2006)
  [hep-ph/0509276];
  %%CITATION = HEP-PH/0509276;%%
  %62 citations counted in INSPIRE as of 19 Mar 2015
%\cite{Nakano:2007dr}
%\bibitem{Nakano:2007dr} 
  E.~Nakano, M.~Nitta and T.~Matsuura,
  ``Non-Abelian strings in high density QCD: Zero modes and interactions,''
  Phys.\ Rev.\ D {\bf 78}, 045002 (2008)
  [arXiv:0708.4096 [hep-ph]];
  %%CITATION = ARXIV:0708.4096;%%
  %57 citations counted in INSPIRE as of 19 Mar 2015
%\cite{Eto:2009kg}
%\bibitem{Eto:2009kg} 
  M.~Eto and M.~Nitta,
  ``Color Magnetic Flux Tubes in Dense QCD,''
  Phys.\ Rev.\ D {\bf 80}, 125007 (2009)
  [arXiv:0907.1278 [hep-ph]];
  %%CITATION = ARXIV:0907.1278;%%
  %41 citations counted in INSPIRE as of 19 mar 2015
%\cite{Eto:2009bh}
%\bibitem{Eto:2009bh} 
  M.~Eto, E.~Nakano and M.~Nitta,
  ``Effective world-sheet theory of color magnetic flux tubes in dense QCD,''
  Phys.\ Rev.\ D {\bf 80}, 125011 (2009)
  [arXiv:0908.4470 [hep-ph]];
  %%CITATION = ARXIV:0908.4470;%%
  %37 citations counted in INSPIRE as of 19 Mar 2015
%\cite{Eto:2009tr}
%\bibitem{Eto:2009tr} 
  M.~Eto, M.~Nitta and N.~Yamamoto,
  ``Instabilities of Non-Abelian Vortices in Dense QCD,''
  Phys.\ Rev.\ Lett.\  {\bf 104}, 161601 (2010)
  [arXiv:0912.1352 [hep-ph]].
  %%CITATION = ARXIV:0912.1352;%%
  %38 citations counted in INSPIRE as of 19 mar 2015


%\cite{Eto:2013hoa}
\bibitem{Eto:2013hoa} 
  M.~Eto, Y.~Hirono, M.~Nitta and S.~Yasui,
  ``Vortices and Other Topological Solitons in Dense Quark Matter,''
  PTEP {\bf 2014}, no. 1, 012D01 (2014)
  [arXiv:1308.1535 [hep-ph]].
  %%CITATION = ARXIV:1308.1535;%%
  %19 citations counted in INSPIRE as of 19 Mar 2015



\bibitem{Abraham:1992vb} 
  E.~R.~C.~Abraham and P.~K.~Townsend,
  ``Q kinks,''  
  Phys.\ Lett.\ B {\bf 291}, 85 (1992);  
  %%CITATION = PHLTA,B291,85;%%
  %\bibitem{Abraham:1992qv} 
  %  E.~R.~C.~Abraham and P.~K.~Townsend,
  ``More on Q kinks: A (1+1)-dimensional analog of dyons,''  
  Phys.\ Lett.\ B {\bf 295}, 225 (1992);  %%CITATION = PHLTA,B295,225;%%
%\bibitem{Arai:2002xa} 
  M.~Arai, M.~Naganuma, M.~Nitta and N.~Sakai,
  ``Manifest supersymmetry for BPS walls in N=2 nonlinear sigma models,''  
  Nucl.\ Phys.\ B {\bf 652}, 35 (2003)  [hep-th/0211103];  
  %%CITATION = HEP-TH/0211103;%%
  %\bibitem{Arai:2003es} 
  %  M.~Arai, M.~Naganuma, M.~Nitta and N.~Sakai,
  ``BPS wall in N=2 SUSY nonlinear sigma model with Eguchi-Hanson
  manifold,''  
  In *Arai, A. (ed.) et al.: A garden of quanta* 299-325
  [hep-th/0302028].  
  %%CITATION = HEP-TH/0302028;%%



%\cite{Isozumi:2004jc}
\bibitem{Isozumi:2004jc} 
  Y.~Isozumi, M.~Nitta, K.~Ohashi and N.~Sakai,
  ``Construction of non-Abelian walls and their complete moduli space,''
  Phys.\ Rev.\ Lett.\  {\bf 93}, 161601 (2004)
  [hep-th/0404198];
  %%CITATION = HEP-TH/0404198;%%
%\cite{Isozumi:2004va}
%\bibitem{Isozumi:2004va} 
  Y.~Isozumi, M.~Nitta, K.~Ohashi and N.~Sakai,
  ``Non-Abelian walls in supersymmetric gauge theories,''
  Phys.\ Rev.\ D {\bf 70}, 125014 (2004)
  [hep-th/0405194];
  %%CITATION = HEP-TH/0405194;%%
%\cite{Eto:2004vy}
%\bibitem{Eto:2004vy} 
  M.~Eto, Y.~Isozumi, M.~Nitta, K.~Ohashi, K.~Ohta and N.~Sakai,
  ``D-brane construction for non-Abelian walls,''
  Phys.\ Rev.\ D {\bf 71}, 125006 (2005)
  [hep-th/0412024];
  %%CITATION = HEP-TH/0412024;%%
%\cite{Eto:2005wf}
%\bibitem{Eto:2005wf} 
  M.~Eto, Y.~Isozumi, M.~Nitta, K.~Ohashi, K.~Ohta, N.~Sakai and Y.~Tachikawa,
  ``Global structure of moduli space for BPS walls,''
  Phys.\ Rev.\ D {\bf 71}, 105009 (2005)
  [hep-th/0503033];
  %%CITATION = HEP-TH/0503033;%%
  %47 citations counted in INSPIRE as of 07 Feb 2015
%\cite{Isozumi:2004vg}
%\bibitem{Isozumi:2004vg} 
  Y.~Isozumi, M.~Nitta, K.~Ohashi and N.~Sakai,
  ``All exact solutions of a 1/4 Bogomol'nyi-Prasad-Sommerfield equation,''
  Phys.\ Rev.\ D {\bf 71}, 065018 (2005)
  [hep-th/0405129].
  %%CITATION = HEP-TH/0405129;%%
  %135 citations counted in INSPIRE as of 09 Dec 2014


%\cite{Manton:1981mp}
\bibitem{Manton:1981mp} 
  N.~S.~Manton,
  ``A Remark on the Scattering of BPS Monopoles,''  Phys.\ Lett.\ B {\bf  110}, 54 (1982).  %%CITATION = PHLTA,B110,54;%%

%\cite{Eto:2006uw}
\bibitem{Eto:2006uw} 
  M.~Eto, Y.~Isozumi, M.~Nitta, K.~Ohashi and N.~Sakai,
  ``Manifestly supersymmetric effective Lagrangians on BPS solitons,''  Phys.\ Rev.\ D {\bf 73}, 125008 (2006)  [hep-th/0602289].  %%CITATION = HEP-TH/0602289;%%


%\cite{Fujimori:2008ee}
\bibitem{Fujimori:2008ee} 
  T.~Fujimori, M.~Nitta, K.~Ohta, N.~Sakai and M.~Yamazaki,
  ``Intersecting Solitons, Amoeba and Tropical Geometry,''
  Phys.\ Rev.\ D {\bf 78}, 105004 (2008)
  [arXiv:0805.1194 [hep-th]].
  %%CITATION = ARXIV:0805.1194;%%
  %26 citations counted in INSPIRE as of 17 Dec 2014

%\cite{Atiyah:1989dq}
\bibitem{Atiyah:1989dq} 
  M.~F.~Atiyah and N.~S.~Manton,
  ``Skyrmions From Instantons,''
  Phys.\ Lett.\ B {\bf 222}, 438 (1989).
  %%CITATION = PHLTA,B222,438;%%
  %154 citations counted in INSPIRE as of 15 mar 2015
%\cite{Atiyah:1992if}
%\bibitem{Atiyah:1992if} 
  M.~F.~Atiyah and N.~S.~Manton,
  ``Geometry and kinematics of two skyrmions,''
  Commun.\ Math.\ Phys.\  {\bf 153}, 391 (1993).
  %%CITATION = CMPHA,153,391;%%
  %43 citations counted in INSPIRE as of 15 mar 2015


%\cite{Harland:2008eu}
\bibitem{Harland:2008eu} 
  D.~Harland and R.~S.~Ward,
  ``Chains of Skyrmions,''
  JHEP {\bf 0812}, 093 (2008)
  [arXiv:0807.3870 [hep-th]].
  %%CITATION = ARXIV:0807.3870;%%
  %10 citations counted in INSPIRE as of 10 mar 2015


%%%%%%%%% discussion

%\cite{Abrikosov:1956sx}
\bibitem{Abrikosov:1956sx} 
  A.~A.~Abrikosov,
  ``On the Magnetic properties of superconductors of the second group,''
  Sov.\ Phys.\ JETP {\bf 5}, 1174 (1957)
  [Zh.\ Eksp.\ Teor.\ Fiz.\  {\bf 32}, 1442 (1957)];
  %%CITATION = SPHJA,5,1174;%%
  %701 citations counted in INSPIRE as of 18 Dec 2014
%\cite{Nielsen:1973cs}
%\bibitem{Nielsen:1973cs} 
  H.~B.~Nielsen and P.~Olesen,
  ``Vortex Line Models for Dual Strings,''
  Nucl.\ Phys.\ B {\bf 61}, 45 (1973).
  %%CITATION = NUPHA,B61,45;%%
  %2020 citations counted in INSPIRE as of 18 Dec 2014

\end{thebibliography}
\end{document}